\documentclass[useAMS,usenatbib]{mn2e}
\usepackage{graphicx}
\usepackage{natbib}  
\usepackage{epstopdf}
\newcommand\aj{{AJ}}%   
\newcommand\araa{{ARA\&A}}%   
\newcommand\apj{{ApJ}}%   
\newcommand\apjl{{ApJ}}%   
\newcommand\aap{{A\&A}}%   
\newcommand\aaps{{A\&AS}}%   
\newcommand\mnras{{MNRAS}}%   
\newcommand\pasp{{PASP}}%   
\def\simgt{\lower.5ex\hbox{$\; \buildrel > \over \sim \;$}}
\def\simlt{\lower.5ex\hbox{$\; \buildrel < \over \sim \;$}}
\newcommand\BMV{B--V} 
\newcommand\VMI{V--I} 
\newcommand\teff{T$_{\rm eff}$}

\newcommand\Teff{T$_{\rm eff}$}

\newcommand{\msun}{\ensuremath{\, {M}_\odot}}
\newcommand{\Msun}{\ensuremath{\, {M}_\odot}}
\newcommand{\ocen}{$\omega$~Cen}
\newcommand{\mlate}{M$_{mixing}$}
\title[]{The fraction of second generation stars in Globular Clusters from the
analysis of the Horizontal Branch}
\author[F. D'Antona and V. Caloi]{F. D'Antona$^{1}$\thanks{E-mail:
dantona@oa-roma.inaf.it (FD); vittoria.caloi@iasf-roma.inaf.it (VC)} and V. Caloi$^2$
$^{1}$\footnotemark[1]\thanks{This work has been supported through PRIN INAF 2005 
``Experimenting stellar nucleosynthesis in clean environments" and 
PRIN MIUR 2007 ``Multiple Stellar Populations in Globular Clusters: 
census, characterization, and origin".
}\\
$^{1}$INAF, Osservatorio Astronomico di Roma, Via Frascati 33, 
00040 Monteporzio Catone (Roma), Italy.\\
$^{2}$ INAF, IASF--Roma, via Fosso del Cavaliere 100, I-00133 Roma, Italy}
\begin{document}

\date{Accepted . Received ; in original form }

\pagerange{\pageref{firstpage}--\pageref{lastpage}} \pubyear{2006}

\maketitle

\label{firstpage}

\begin{abstract}
The majority of Globular Clusters show chemical inhomogeneities in the
composition of their stars, apparently due to a second stellar generation in
which the forming gas is enriched by hot-CNO cycled material processed in
stars belonging to a first stellar generation. Clearly this evidence prompts questions on
the modalities of formation of Globular Clusters. An important preliminary
input to any model for the formation of multiple generations is to determine which
is today the
relative number fraction of ``normal" and anomalous stars in each cluster.  As
it is very difficult to gather very large spectroscopic samples of Globular
Cluster stars to achieve this result with good statistical significance, we
propose to use the horizontal branch.
We assume that, whichever the progenitors of the second generation, the
anomalies also include enhanced helium abundance. In fact, helium variations
have been recently recognized to be able to explain several puzzling
peculiarities (gaps, RR Lyr periods and period distribution, ratio of blue to
red stars, blue tails) in horizontal branches.  We summarize previous results
and extend the analysis in order to infer the percentage in number of the first and
second generation in as many clusters as possible. We show that, with few exceptions,
approximately 50\% or more of the stars belong to the second generation. In
other cases, in which at first sight one would think of a simple stellar
population, we give arguments and suggest that the stars might all belong to the second
generation. 
We provide in Appendix a detailed discussion and new fits of the optical and UV data
of NGC~2808, the classic example of a multiple helium populations cluster, 
consistently including a reproduction of the main sequence splittings
and an examination of the problem of ``blue hook" stars. We also show a detailed fit 
of the totally blue HB of
M~13, one among the clusters that are possibly fully made up by second generation stars.
We conclude that the formation of the second generation is a crucial event in the life
of globular clusters. The problem of the initial mass function
required to achieve the observed high fraction of second generation stars can
be solved only if the initial cluster was much more massive than the present one
and most of the first generation low mass stars have been preferentially lost. 
As shown by D'Ercole et al. by modelling the formation and dynamical evolution
of the second generation, the mass loss due to the explosions of the type II supernovae 
of the first generation may be the process responsible for triggering the expansion of the
cluster, the stripping of its outer layers and the loss of most of the
first generation low-mass stars.
\end{abstract}


\section{Introduction}
\label{sec:intro}

The observations of Globular Cluster (GC) stars are still to be interpreted in
a fully consistent frame.  Nevertheless, a general consensus is emerging on
the fact that most GCs can not be considered any longer ``simple stellar
populations" (SSP), and that ``self--enrichment" is a common feature among
GCs. This consensus follows from the well known ``chemical anomalies'',
already noted in the seventies (such as the variations found in C and N abundances,
the Na--O and Mg--Al
anticorrelations). Recently observed to be present at the turnoff (TO) and
among the subgiants \citep[e.g.][]{gratton2001,briley2002, briley2004, cohen2005}, they
must be attributed to some process of ``self--enrichment" occurring at the
first stages of the cluster life,
%INSERT REFEREE
as the same authors quoted above suggest. 
%END INSERT
There was a first epoch of star
formation that gave origin to the ``normal" (first generation, hereinafter FG)
stars, with CNO and other abundances similar to Population II field stars of
the same metallicity. Afterwards, there must have been some other epoch of
star formation (second generation, hereinafter SG), including material heavily
processed through the CNO cycle.  This material either comes entirely from the
stars belonging to the first stellar generation, or it is a mixture of processed
gas and pristine matter of the initial star forming cloud. We can derive this
conclusion as a consequence of the fact that there is no appreciable
difference in the abundance of elements such as Ca and the heavier ones
between ``normal" and chemically anomalous stars belonging to the same
GC. Needless to say, this statement {\it does not} hold for $\omega$ Cen,
which must indeed be considered a small galaxy and not a typical GC. In the
following, we will only examine ``normal clusters", those which do not show
signs of metal enrichment due to supernova ejecta.  The homogeneity in the
heavy elements is an important fact that tells us, e.g., that it is highly
improbable that the chemical anomalies are due to mixing of stars born in two
different clouds, as there is no reason why the two clouds should have a
unique metallicity. In addition, the clusters showing chemical anomalies have
a large variety in metallicities, making the suggestion of mixing of two
different clouds even more improbable.  The matter must have been processed
through the hot CNO cycle, and not, or only marginally, through the helium
burning phases, since the sum of CNO elements is the same in the ``normal" and
in the anomalous stars \citep[e.g.][]{smith1996,ivans1999,cohenmelendez2005}. 
\cite{carretta2005} find that
actually the CNO is somewhat --but not much-- larger in the SG
stars of some GCs. Therefore, the progenitors
may be either massive asymptotic giant branch (AGB) stars \citep[e.g.][]{ventura2001,
  ventura2002}\footnote{If the \cite{carretta2005} CNO data really
  indicate that a limited number of third dredge up \citep[e.g.][]{ibenrenzini1983}
  episodes plays a (small) role in the nuclear processing of the matter giving origin to the SG, 
the massive AGB progenitors are possibly
favoured.} or fast rotating massive stars \citep{decressin2007}.  In both cases, models
show that the ejected material must be enriched in helium with respect to the
pristine one. The higher helium content has been recognized to have
a strong effect on horizontal branch (HB) morphology, possibly helping to
explain some features (gaps, hot blue tails, second parameter) which,
until now, have defied explanation \citep{dantona2002}. Along these lines, a
variety of problems has been examined: the extreme peculiarity of the HB
morphology in the massive cluster NGC~2808, \citep{dantonacaloi2004}; the second
parameter effect in M~13 and M~3 \citep{caloi-dantona2005}; the peculiar
features in the RR Lyr variables and HB of NGC~6441 and NGC~6388
\citep{caloidantona2007}. The presence of strongly enhanced helium in peculiar
HB stars has been confirmed, for NGC 2808 and NGC 6441, by spectroscopic observations
\citep{moeler2004,busso2007}.

Beside this spectroscopic evidence, an unexpected feature has recently
appeared from photometric data: the splitting of the main sequence in NGC
2808.  After first indications from a wider than expected colour distribution
\citep{dantona2005}, recent HST observations by \cite{piotto2007} leave no
doubt that there are at least three different populations in this
cluster. This came after the first discovery of a peculiar blue main sequence
in $\omega$~Cen \citep{bedin2004}, interpreted again in terms of a very high
helium content \citep{norris2004, piotto2005}.
The above mentioned cases can be considered as ``extreme'' ones, in the sense
that no explanation had been attempted for them before the hypothesis of
helium-enriched populations. Less critical situations, such as the HB
bimodality in NGC 1851 and 6229, had been tentatively explained in terms of a
unimodal mass distribution with a large mass dispersion (0.055 -- 0.10 \msun,
Catelan et al. 1998). But even such a rather artificial assumption could not
help in the case of NGC 2808, which, as hinted before, finds a possible solution 
only in terms of varying helium in multiple stellar generations.
Therefore, we consider appropriate to apply to the less peculiar cases
the solution found plausible for the most peculiar ones. In fact, notice that
split main sequences and strong bimodalities are only the tip of 
the iceberg of the self--enrichment phenomenon. In most clusters the higher
helium abundances remain confined below Y$\sim 0.30$, and the presence of such
stars would not be put in evidence either from main sequence observations
\citep{dantona2002,salaris2006}, or from a naif interpretation of stellar
counts on the HB, as we shall discuss in Sect. \ref{ngc6441}.
 
If we wish to shed light on the entire process of
formation of GCs we must have a rough idea of the total number of SG stars.
We will then analyze the HB in terms of populations differing in Y, using one or
more of the following peculiar features:
\begin{enumerate}
\item bimodal HBs and HB gaps;
\item presence of blue HB stars and very long period RR Lyr's in high Z clusters;
\item peaked number vs. period distribution of RR Lyr's;
\item blue--HB clusters. 
%\item presence --or not-- of extreme HB (EHB) stars and ``blue hook" stars 
%\cite{sweigart1997,brown2001,cassisi2003}
\end{enumerate}
In this paper we show the results of such an interpretation for several GCs. We start 
from a reanalyis of the NGC~2808 data, taking into account the results by 
\cite{piotto2007} for the main sequence, and the ultraviolet HST data by 
\cite{castellani2006}; we summarize the results already published and discuss 
briefly the other clusters. The table of the derived FG and SG percentages is 
the basis to discuss the clusters' dynamical evolution required to produce the 
high fraction of stars presently belonging to the SG.

\section{The basic model}\label{sec:2}

Here we summarize why and how a helium enrichment modifies the HB morphology 
\citep{dantona2002}, and the basic inputs of the HB and main sequence (MS) simulations adopted to 
constrain the FG and SG. 

\subsection{Red giant mass, mass loss, and helium content}
\label{sec:2.1}
The evolving mass in a GC is a function of age, metallicity and helium content. 
From the well defined turnoff of GCs, it is evident that any age spread must be 
much smaller than the global age (10-13 Gyr). In addition, the metallicity 
spread among cluster members are contained within $\sim 0.04$dex \footnote{ 
According to \cite{gratton-ar}: ``Low upper limits in the spread of the 
abundances of Fe (of the order of 0.04 dex, r.m.s.) have been found for several 
clusters from both spectroscopy and from the widths of main sequence (MS) and 
red giant branch (RGB) stars in the colour magnitude diagrams CMDs (for summary 
and discussion, see Suntzeff 1993). At present the verdict on Fe 
variations in CGs except \ocen\ must remain `not proven'."}. Thus, once fixed 
age and metallicity, the evolving mass is a function of the helium content 
only, and it decreases when helium is increased. The dependence $\delta 
M_{RG}/\delta Y \sim -1.3 \msun$\ is such that, e.g. the small increase in 
helium content from the primordial value Y=0.24 to the moderately higher Y=0.28 
decreases the evolving mass by $\sim$0.05\msun. 

During the latest phases of red 
giant evolution, both normal and helium enhanced stars lose mass.
We assume that mass loss follows Reimers' law \citep{reimers1975}
\begin{equation}
    \dot M_R=4 \cdot 10^{-13} \eta_R {LR\over M}  
\end{equation}  
where $\eta_R$ is a free parameter directly connected with the mass loss rate 
and L, R and M are luminosity, mass and radius expressed in solar units. 
%INSERT REFEREE
This expression has no explicit dependence on helium abundance or 
metallicity. While the independence from the metallicity may be questioned, 
the helium content, at the level of variation we are considering,
should not affect the mass loss rate, as helium 
has no strong effect on surface opacities and possible grain formation.
%END REFEREE
It turns out that also the total mass lost by giants with different helium content and 
similar age does not depend on the helium content. In 
fact, we computed tracks of stars with different helium content, and having 
different mass so that they have the same evolving age, and we find that {\it 
the total mass lost} differs only by 0.001 -- 0.002\msun when the models reach 
the helium flash. On the other hand, the helium core mass at the helium flash 
depends on the helium content, decreasing with increasing helium, but at a much 
smaller rate than the decrease of the red giant mass. Therefore, the ratio of 
core mass to the remnant mass is larger for larger Y, and these stars will 
occupy a position on the ZAHB at a bluer colour than stars with lower Y. This 
means that, if the cluster contains stars of FG with ``standard" Y, and stars 
of SG with larger Y, these latter will be ``bluer" than the FG stars.

Assuming the point of view that the HB contains stars with different helium 
content, the estimate of age and metallicity on the one side, and the HB 
morphology on the other, will indicate which part of the HB population belongs 
to the FG. This component should have a uniform helium content, probably close 
to the Big Bang abundance \citep[e.g. Y$\sim$0.24][]{coc2004}. For the FG of the most 
metal rich clusters we will assume a larger initial, uniform helium Y=0.25. On 
the contrary, the SG will most probably show a spread in Y, for two main 
reasons: the self--enriched material in fact 1) may come from different 
progenitors, having different chemical peculiarities or 2) it may be diluted in 
different fractions with matter from the FG. In both cases the helium abundance 
may differ among the SG stars.  Notice however that {\it if we have physical 
reasons (e.g. based on one of the peculiarities listed in Sect. 
\ref{sec:intro}) to attribute to a star a helium content larger than the helium 
of the FG, we know that this star belongs to a SG, even if its Y is not much 
larger.} Of course, the total amount of SG self--enriched gas differs if the 
derived Y does, or does not, result from dilution with pristine matter.

\subsection{The grids of HB models}
\label{sec:hb}
The basis of the synthetic HB distributions are stellar models computed with 
the code ATON2.0,  described in \cite{ventura1998}  and \cite{mazzitelli1999}. 
The HB models have been evolved until the disappearance of helium in the 
convective core. We have adopted metallicity Z=2 $10^{-4}$ for the metal poor 
GCs; Z=$10^{-3}$ and Z=2$\times 10^{-3}$\ for the intermediate metallicity 
clusters, which represent the majority of data sets, and Z=6$\times 10^{-3}$\ 
for the high metallicity clusters. The helium core mass of the models
is set at the helium flash core mass of the previous evolution,
determined by evolving models for each couple (Z, Y) for an age of
11$\times 10^9$yr. The metallicities of 
individual clusters may be slightly different from those adopted here, but the main aim 
---to distinguish between normal--helium and enhanced--helium stars in the 
construction of the HB--- can be satisfactorily achieved. 
We will see that the ratios FG/SG are well defined from gross characteristics 
of the HB, and not by minute details. We computed grids of models for Y=0.24, 
0.28, 0.32 and 0.40.

We computed HB models up to \Teff$\sim$31000K, that is the usually accepted
limit for standard ZAHB models resulting from a He--flash occurring at the
red giant branch tip.

These models can not explain the extreme ``blue hook" stars (T$_{\rm eff}$ up to $\sim$37000K)
present in   $\omega$Cen, M54 \citep{rosenberg2004}, NGC6388 \citep{busso2007}
and NGC 2808 \citep{moeler2004}, and generally explained as a result of the mixing
of processed matter with the very small residual hydrogen envelope,
consequent to a late ignition of the helium flash, along the white dwarf cooling sequence
 \citep{sweigart1997, brown2001}. Mixing raises the helium
 (and carbon) abundance in the envelope \citep[see also][]{cassisi2003} and the
 star settles at larger \Teff. 
In addition, \cite{dantona2007} have suggested that deep mixing occurs, independently from
a late helium flash, along the RGB evolution of giants belonging to the very high helium population
(Y$\sim$0.35 -- 0.40) as deep mixing in this case is not forbidden by the molecular weight
discontinuity barrier. So, deep mixing may involve most of the very helium rich
stars, and increase, even considerably, their surface helium abundance\footnote{The hypothesis
of deep mixing is useful to increase the probability of obtaining extreme HB and blue hook stars. If
we use only the very late flash hypothesis, just a few stars could happen to have the
appropriate envelope mass, such that they do not ignite the flash on the RGB, or do not leave a He--white
dwarf remnant.}.

In order to simulate the possible late--flash induced mixing (and any other kind of deep mixing),
for the set of Z=0.001 adopted to simulate the HB of NGC~2808,   
we computed tracks of very low masses for Y=0.45, 0.50, 0.60, 0.70 
and 0.80, having core mass M=0.4676\Msun, the helium core mass at the flash for the Y=0.40 models.  

\subsection{Main sequence and turnoff models}
For comparison with the main sequence colour distribution in NGC~2808 by 
\cite{piotto2007}, we use the isochrones for Z=$10^{-3}$ and Y=0.24, 0.28, 0.32 
and 0.40 described in  \cite{dantona2005}. 

%content is approximated by: 
%\begin{eqnarray} 
%\log M_{RG}/M_{\odot} \simeq [-0.282+0.092(Y-0.24)]\times \log t   \nonumber \\ 
%    -1.693(Y-0.24)+2.768+12(Z-10^{-3}) 
%\label{eq1} %\end{eqnarray} 

\subsection{Simulations}
\label{sec:simul}
We adopt an appropriate relation between the mass of the evolving giant
$M_{RG}$ and the age, as function of helium content and metallicity. 
The mass on the HB is then: 
\begin{equation} 
M_{HB} =  M_{RG}(Y,Z) - \Delta M \label{eq2} 
\end{equation} 
$\Delta M$\ is the mass lost during the RG phase. We assume that $\Delta M$\ 
has a gaussian dispersion $\sigma$\ around an average value $\Delta M_0$\  and 
that both $\Delta M_0$\  and $\sigma$\ are parameters to be determined and do 
not depend on Y.\footnote{ A complication in the interpretation of HB 
morphologies, not included in this work, arises if not only the helium 
content between FG and SG varies, but {\it also the total CNO content}. This 
additional problem comes out from the observations of NGC~1851 (see Sect. 3.2) 
whose HR diagram shows a splitting in the subgiant branch \citep{milone2008}. 
While \cite{cassisi2008} are able to explain this feature by assuming an SG 
with about double CNO content, and same age as the FG (thus in the classical 
self--enrichment scenario), the subsequent interpretation of the HB morphology 
by \cite{salaris2008} can not be simply achieved by doubling the CNO, but it 
requires some extra mass loss for the SG stars. Here, our main aim is to find out the 
percentages of FG and SG, and this can be obtained also within the framework of 
our simpler assumptions.} Therefore the HB mass varies both due to the mass 
dispersion $\sigma$\ around the average assumed mass loss, and due to the 
dependence of the RG mass on the helium content. As the evolving mass {\it 
decreases} with increasing helium content, the stars with higher helium will 
populate bluer regions of the HB. In each cluster, we must identify the FG 
section of the HB (e.g., in NGC 2808 this is easily identified with the red 
clump). Then we assume a primordial (Y=0.24) or a minimum (Y=0.25) helium content 
for the fraction of HB stars considered to belong to the FG.  This fraction can be 
adjusted in order to reproduce the features we attribute to the FG (e.g., the 
peaked RR Lyr period distribution in M3, see later). We then assume that the 
rest of HB stars has larger Y, and adopt different N(Y) distributions in order 
to reproduce the other parts of the HB number vs. colour distributions, when
available.

We fix $\Delta M_0$\ and $\sigma$, and extract random both the mass loss and 
the HB age in the interval from 10$^6$yr to 10$^8$yr, according to the chosen Y 
distribution. We thus locate the luminosity and \Teff\ along the evolution of 
the HB mass obtained.  We identify the variable stars as belonging to a fixed 
\Teff\ interval and compute their period according to the pulsation equation 
(1) by \cite{dmc2004}. The results are very similar if we adopt the classic 
\cite{va1973} relation. The real problem is given by the choice of the exact 
boundaries of the RR Lyr strip, that affect strongly the number and mean period 
of the RR Lyrae variables \citep[see, e.g., the discussion 
in][]{caloidantona2008}.  

The L and \teff\ values are transformed into the different observational bands. 
As most observations are available in the B and V bands, and/or in the 
Bessell's I, we derive the visual magnitude M$_{\rm v}$ and the \BMV\ or \VMI\ 
colours, using the transformations by \cite{bessell-castelli-plez1998}. Although our main aim ---to 
understand the different fractions of FG and SG stars--- is not affected by this 
problem, we remark that the \BMV\ and \VMI\ colours saturate at large \teff. The 
bolometric corrections become very large, so the number vs. magnitude 
distribution may suffer some uncertainties, which can be avoided by using 
different magnitudes, e.g. the ultraviolet HST bands. We exemplify such 
comparisons for the case of the clusters NGC~2808 and M~13 in the Appendix A. 
The transformations for the ACS -- HST bands are taken from \cite{bedincastelli2005}.
The WFPC2--HST relations are by \cite{origlia2000}, plus additional 
transformations kindly provided by L. Origlia.
 
We associate a gaussian spread in colour and magnitude to each point, in order 
to simulate the impact of observational errors.  We do not include binaries in 
the simulations.  We mostly compare the theoretical simulations and the 
observations by looking at the number counts vs. colour in the horizontal part 
of the HB (e.g. for the red part and the RR Lyr) or vs. magnitude, for the 
vertical part, if the blue HB is very extended. 

\section{Bimodality and gaps in the HB}
\label{sec:bimodalhb}
There is a huge literature which has defined and attemped to understand the gaps 
on the blue side of the HB \citep[see, e.g. ][and references therein]{ferraro1998, piotto1999}. 
While a gap at \teff$\sim 10^4$K should probably be attributed to the operation 
of diffusion \citep{glaspey1989, grundahl1999,caloi1999}, other gaps may have to do 
with discontinuities in the helium content \citep{dantona2005,lee2005}.  In particular, a 
bimodal distribution in colour characterizes the HB of some clusters, that have well populated blue
 and red sides of the HB, with limited or null population of RR Lyr variables. NGC~2808 is a 
prototype of this class, and \cite{catelan1998} were not able to interpret it in 
terms of a unique mass distribution, even with a mass spread as large as 0.3\msun. 
A multimodal mass distribution was then required to explain this HB.
  
The HB bimodality in NGC~2808 was the main hint used by \cite{dantonacaloi2004} 
to infer the presence of multiple stellar generations differing in Y in the 
cluster, an interpretation nicely supported by the subsequent observation of 
the main sequence splitting \citep{dantona2005,  piotto2007}.  In addition, in 
this cluster are found the still misterious ``blue hook" stars. So NGC~2808 
appears as an ideal benchmark for the application of the multiple population 
hypothesis. We summarize in the following our present understanding 
of its modeling, and extend it to other situations.

\subsection{NGC 2808}

If age, metallicity and mass loss are such that normal--helium, FG stars, 
populate a red clump, the SG stars with helium enhancement will tend to 
populate the bluer HB and the RR Lyr region. If there is a gap between the 
normal--helium stars and the {\it minimum} helium content of the second 
generation, the case of NGC~2808 may appear: a red clump (FG), almost no RR Lyr 
(due to the helium gap) and a blue HB with larger helium content \citep[starting from 
Y$\sim 0.28$\ according to ][]{dantonacaloi2004, dantona2005}.  In those 
papers, we had modelled the mass loss along the RG branch by assuming that the 
larger Y would provoke a slightly larger global mass loss, as the evolving 
giants with higher Y are less massive, and thus have smaller gravity 
\citep[see, e.g.][]{lee1994}. We have ascertained that this is not the case 
(Sect. \ref{sec:2.1}), so the HB has to be modeled by assuming the same average 
mass loss for both normal Y and higher Y red giants. In addition, we try to 
model better the EBT2 and EBT3 \citep[in the definition by][]{bedin2004} blue 
clumps, which contain the extreme HB \citep{dcruz2000,brown2001} and the 
``blue hook" stars \citep{moeler2004}, respectively, and compare also 
simulations based on HST ultraviolet and visual bands. The details can be found 
in the Appendix. We show that the new simulations are consistent with both the 
triple MS by \cite{piotto2007} and the HB star distribution. The main 
difference with respect to the analysis by \cite{dantona2005} is that the 
intermediate Y population is now clustered at Y$\sim$0.31.  The cluster again 
results divided into 50\% normal--helium stars, and 50\% helium enriched stars, 
although the very high helium (Y$\sim$0.385) stars are only $\sim$15\%.

\subsection{NGC 1851, NGC 6229}
 
Following the guidelines mentioned in Sect. \ref{sec:intro}, we take the point 
of view of interpreting clusters with a bimodal HB in terms of multiple 
populations, as in the ``paradigmatic'' case of NGC~2808. \cite{catelan1998} 
and \cite{borissova1999} describe the complex HB structure of the 
cluster NGC~1851 and NGC~6229, respectively: bimodal, with few RR Lyr variables, a gap on the 
blue HB and possibly some extreme blue HB members, at the luminosity of the 
turn--off. In NGC~6229, the number ratio of the red, variable and blue HB 
components are: B:V:R=0.59:0.08:0.33.  The RR Lyr average period is of the OoI 
type. We assume therefore that the RR variables belong to the first generation, 
since they have the period appropriate to the metal and helium content expected 
for these stars. So we expect that roughly 41\% of cluster members belong to 
the first generation and 59\% to the second, helium--enriched one. The extra--
helium allows these latter stars to reach the bluer positions beyond the 
variable region, to which they should have been confined by a chemical 
composition of Y$\sim$0.24 and Z$\sim$0.001. (See later the case of M3). Of course, not 
all the helium--enhanced stars will be found on the blue, and vice--versa, but 
the distribution will be substantially the one mentioned above.

Similarly, for NGC~1851 we have: B:V:R:=0.30:0.10:0.60 \citep{catelan1998}, or 
B:V:R$\sim$0.32;0.12;0.56 \citep{walker1998}. In this case the RR~Lyr are again 
Oosterhoff type I, but their periods are longer, so that the variables may 
belong to the SG. Recently, this latter cluster has been discovered to harbour 
a double subgiant branch \citep{milone2008}, that can be interpreted as the 
presence of an FG and SG with different total CNO abundances and same age 
\citep{cassisi2008}. The ``bright" subgiant branch contains 55$\pm$5\% of stars 
\citep{milone2008} and may correspond to the red part of the HB. Thus the FG 
should contain $\sim$55\% of the total cluster stars. As we already remarked, 
\cite{salaris2008} analyze the HB stellar distribution, and are not able to fit 
the blue and red side of the HB with different helium (and CNO-Na abundances) 
and the same mass loss on the RGB. While modelling of this HB may require a 
database of HB tracks computed with different Y and  different compositions 
in CNO, for our present purposes it is sufficient to see that the 
consistent bimodality of both the SGB and HB indicate the presence of an FG and 
SG, with the given proportions.

\section{Clusters with high metallicity and peculiar HB}
\subsection{NGC 6441 and NGC 6388}
\label{ngc6441}

The case of the high metallicity cluster NGC~6441 has been fully discussed by 
\cite{caloidantona2007}, who showed that very helium rich stars {\it are also 
present among the red clump stars}. The analysis is able to explain not only 
the anomalous long periods of the RR Lyr \citep{pritzl2003}, but also the 
extension in magnitude of the red clump (any attempt to attribute this 
thickness to differential reddening has failed ---\cite{raimondo2002}), and the 
hot blue side of the HB. Therefore, the morphology requires  helium enrichment 
not only for the bluer side of the HB, as we could naively think, but even 
for the red clump stars.  The physical reasons for this 
interpretation is the following: helium core burning stars having high Y and Z 
make long loops from red to blue in the HB \citep{sweigart-gross1976}. In fact, 
both the higher mean molecular weight --leading to a high H--burning shell 
temperature-- and the high metallicity --leading to a stronger CNO shell--
conspire towards the result that the H--shell energy source prevails with 
respect to the He--core burning.  The consequent growth of the helium core 
leads the evolution towards the blue.  Therefore, if we must explain the luminous 
(long period) RR Lyr by stars having high helium, the same stars will also 
populate the red clump: this is exactly what we observe: if the helium content 
is not as large as Y$\sim 0.35$\ {\it in the red clump}, the HB finds no 
satisfactory explanation. The percentage of helium enriched stars is in this 
case $\sim$60\% for NGC 6441.

We performed a similar analysis for NGC~6388 (see Fig.~1). The main 
difference among the two clusters is that NGC~6388 seems to have a higher tail 
of very high helium (Y$>$0.35) stars, reaching $\sim$20\%. 
\begin{figure} 
\centering \includegraphics[width=8cm]{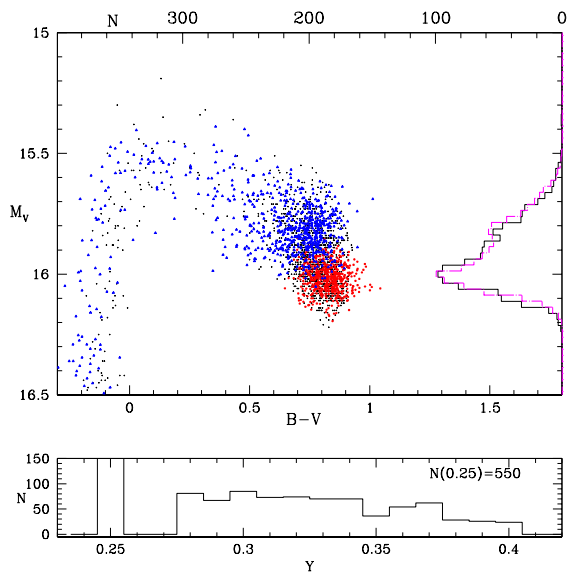} 
\caption{Synthetic HB simulation for NGC~6388. A similar analysis for NGC~6441 
is shown in Caloi \& D'Antona (2007). The HB data are taken from Piotto et al. (2002). 
In black we have the observations, while red and blue are the simulated stars.
In red are the Y=0.25 stars, in blue those with Y$>$0.25 for a total of 1300 stars.
The observed clump distribution (full line histogram) is shown on the right, 
superimposed to the theoretical distribution (dash--dotted 
histogram). The bottom panel shows the number vs. helium 
distributions assumed for the simulation. The number of stars with primordial 
helium Y=0.25, N(0.25), is indicated in the label.} 
\label{fig1} 
\end{figure} 

There are other analyses for these clusters in the literature: \cite{busso2007} 
consider as peculiar only the blue HB stars. This would limit the SG to $\sim 
15$\%. Also \cite{yoon2008} attribute the presence of a SG to the RR Lyr 
and hotter stars only. They seem to be able to obtain the RR Lyr long periods with 
only Y$\sim$0.3, but we have no details about their models to understand this 
difference. Our models require Y up to $\sim$0.35 to fit the long periods of 
the RR Lyr, but, as explained above, this high Y also helps to reproduce a 
thickness $>0.7$mag of the red clump.  On the other hand, \cite{yoon2008} 
attribute the thickness of the red clump to differential reddening, a
hypothesis in contrast with the data analysis by \cite{raimondo2002}.
 
\subsection{47 Tuc}
In 47~Tuc, the prototype of metallic GCs, the red clump is much less thick in 
magnitude than in the two anomalous clusters above. In Fig.~2 we show the 
histograms of the number of stars in the red clump of NGC~6441, NGC~6388 and 
47~Tuc versus magnitude. The magnitudes have been normalized so that the peak 
in the distribution coincides for all the clusters. The ``thickness" in 
magnitude of NGC 6388 and NGC 6441 is larger than for 47 Tuc. The excess 
of stars at smaller luminosities below the maximum is probably due to the 
larger observational errors, but the (asymmetric) excess at higher luminosities 
is most easily interpreted as due to stars with helium much higher than normal. 
Based on this feature only, we infer that the 47~Tuc SG should not be larger 
than $\sim$25\% of the stars.  Observations show that CN strong and CN weak 
stars in the cluster  are about in similar percentages \citep{briley2004}, and 
if these two groups are to be interpreted as FG and SG, we have a 
contradiction. An escape from this problem can be found if the first stellar 
generation in 47~Tuc has a larger initial helium content \citep{salaris1998}.

\begin{figure}
   \centering
   \includegraphics[width=8cm]{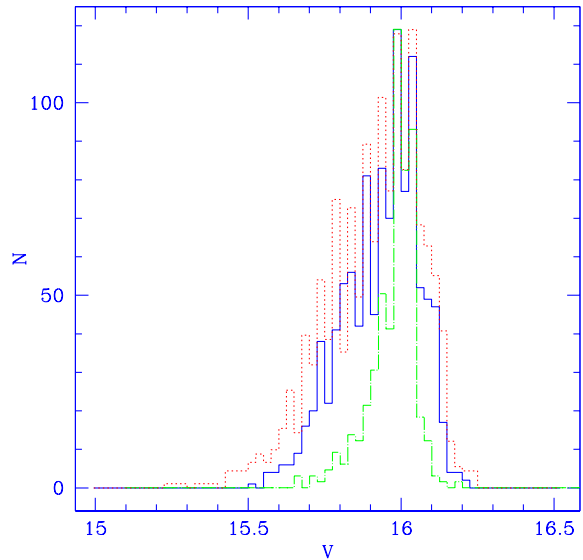}
      \caption{Plot of the observed number of stars versus magnitude for the red horizontal branch
	  of three metal rich clusters: NGC 6441 (dots), NGC 6388 (full line), and 47 Tuc 
	  (dashed line). See text for the interpetation.
	}
         \label{fig2}
   \end{figure}

\begin{table*}
\caption{Helium history of clusters}
\smallskip
\begin{center}
{\small
\begin{tabular}{l|cc|cc|cc|cc}
\hline
\noalign{\smallskip}
Name & \multicolumn{2}{c}{FG}& \multicolumn{2}{c}{SG}& \multicolumn{2}{c}{Extreme pop.}
& Data & Interpretation \\\hline
  & Y	 & \% &  Y	 & \% &  Y	 & \% &&  \\
\noalign{\smallskip}
\hline  
\noalign{\smallskip}
\noalign{\smallskip}
\multicolumn{5}{l}{Bimodal HB (and gaps). Blue MS} &&&&\\
\noalign{\smallskip}
\hline
$\omega$Cen & 0.24 & ?   &   &  ?   &  $\sim$0.38 &  $\sim$20 -- 25   &  1  &  2    \\
NGC 2808 & 0.24	& 50 &	0.30-0.32 &	35& $\sim$0.38 & 15      & 3, 4 & 5, 6, this paper  \\
NGC 1851 & 0.24 & 65 &      ?     & 35 &&& 7, 8 & 9, this paper\\
NGC 6229 & 0.24 & 40 &  $>$0.30 & 60 &&& 7, 10 & this paper\\
\hline
\noalign{\smallskip}
\noalign{\smallskip}
\multicolumn{5}{l}{High Z -- Anomalous HB} &&&   &\\
\noalign{\smallskip}
\hline
NGC 6441 & 0.25 & 38 & 0.27-0.35   &48&	$>$0.35 &14 &  11, 12, 14  & 13, 14 \\
NGC 6388 & 0.25	& 39 & 0.27-0.35&	41 & $>$0.35 &	20 &  11, 14, 15 & 14, this paper\\
47 Tuc & 0.25	&75 (?)  & 0.27-.32	&25 (?)  && & 11 & this paper \\
47 Tuc & 0.27	&50 (?)  & 0.29-.32	&50 (?)  && & 16 & 17, this paper \\
\hline
\noalign{\smallskip}
\noalign{\smallskip}
\multicolumn{5}{l}{Peaked distribution of RR~Lyr's periods} &&& \\
\noalign{\smallskip}
\hline
M3     & 0.24 & 50 & .26--.28 & 50 &&& 18, 19  & 20 \\
M5     & 0.24 & 30 & .26--.31 & 70 &&& 21   & this paper \\
NGC 3201& 0.24 & 63 & .26 - 0.28 & 37 &&& 22 & this paper\\
NGC 7006& 0.24 & 72 & 0.25 - 0.275 & 28 &&& 23  & this paper\\
M68     & 0.24 & 45 & 0.26 - 0.28   & 55 &&& 24 & this paper\\
M15     & 0.24 & 20 & 0.26 - 0.30   & 80 &&& 25 & this paper\\ 
\hline
\noalign{\smallskip}
\noalign{\smallskip}
\multicolumn{5}{l}{ Blue--HB clusters } &&& \\
\noalign{\smallskip}
\hline
M53    & 0.24 &  0? & 0.27--0.29? & 100 &&& 26, 27 & this paper\\
M13    & 0.24 & 0? & 0.27--0.35  & 70 & $\sim$0.38  & 30 & 28  & 29, this paper\\
NGC 6397 & 0.24 & 0? & 0.28 (?) & 100 &  && 30, 31, 32  & this paper\\
\noalign{\smallskip}
\hline
\end{tabular}}
\end{center}
\medskip
(1)\cite{bedin2004}, \cite{piotto2005}; (2) \cite{norris2004}; (3) \cite{bedin2000,castellani2006}; 
(4) \cite{piotto2007}; (5) \cite{dantonacaloi2004, dantona2005}; (6) \cite{lee2005}; 
(7) \cite{catelan1998, walker1998}; (8) \cite{milone2008}; (9) \cite{cassisi2008, salaris2008};  
(10)  \cite{borissova1999}; (11) \cite{piotto2002}; (12) \cite{pritzl2003}; 
(13) \cite{caloidantona2007}; (14) \cite{busso2007}; (15) \cite{pritzl2002};
 (16) \cite{briley2004}; (17) \cite{salaris1998}; 
(18) \cite{cc2001}; (19) \cite{ferraro1997, buonanno1994}; (20) \cite{caloidantona2008}
(21) \cite{sandquist2004} (22) \cite{layden2003, piersimoni2002} 
(23) \cite{wehlau1999}; (24) \cite{walker1994}; (25) \cite{clement2001}; 
(26) \cite{rey1998}; (27) \cite{martell2008}; (28) \cite{ferraro1998}; 
(29) \cite{caloi-dantona2005}; (30) \cite{kaluzny1997}; (31) \cite{king1998,richer2006}; (32) \cite{carretta2005,
bonifacio2002, pasquini2008}
\cite{caloidantona2008}
\label{table1}
\end{table*}
 
\section{The peaked period distribution of RR Lyr stars}

The prototype of HBs has often been considered the HB of the cluster M3: it is
well populated both in the red part, in the RR Lyr's and in the blue side,
without a blue tail. The HB distribution among red, variable and blue members
can be reproduced by assuming an average mass loss along the RGB, with a
standard deviation $\sigma \sim 0.025$\Msun.

\begin{figure*}
   \centering
   \includegraphics[width=8cm]{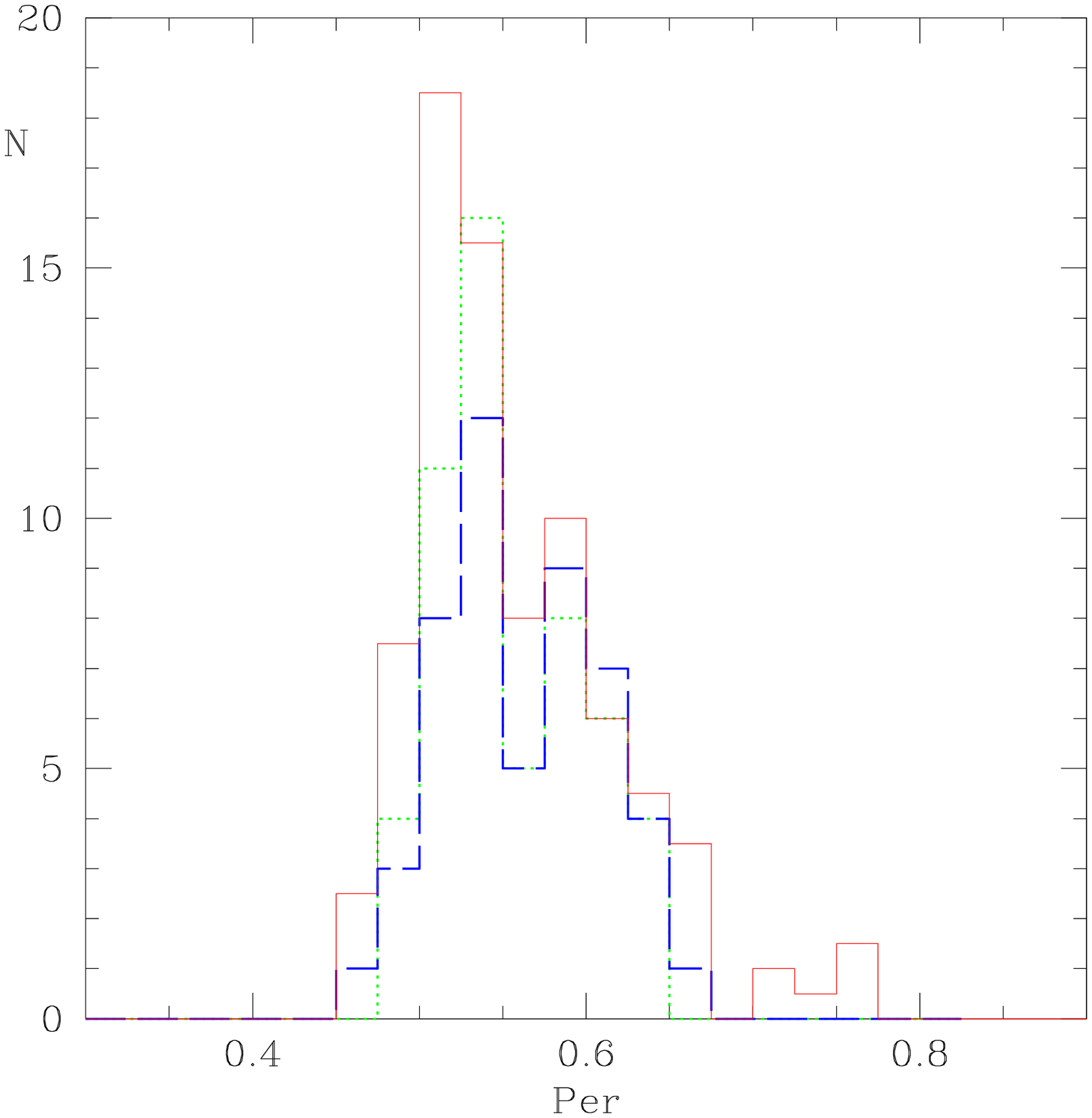}
   \includegraphics[width=8cm]{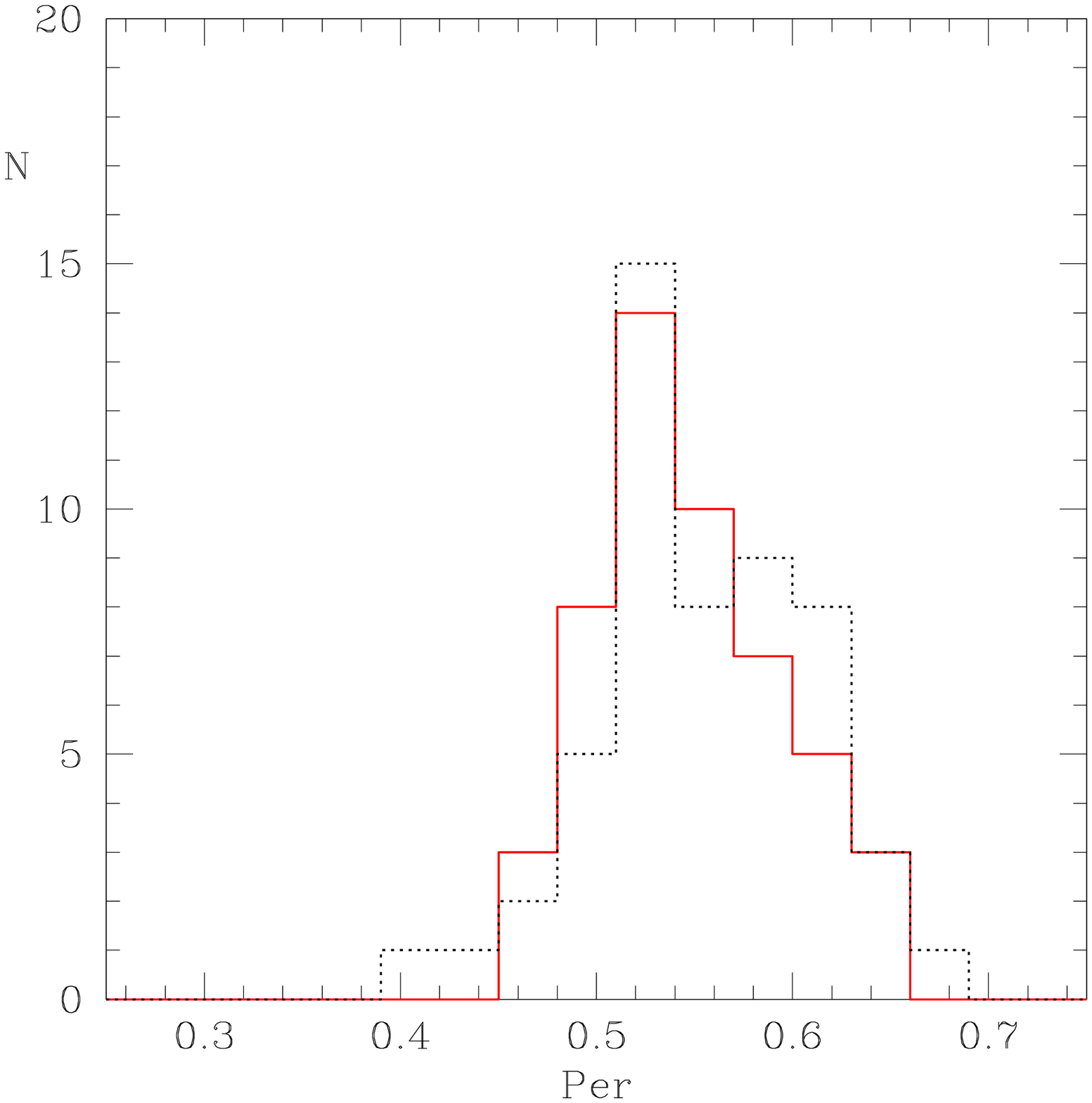}
      \caption{In the left figure we plot the number of RR Lyr vs. pulsation period for NGC~3201 
	  from two different databases: dashed line, by Layden et al. 2003, 
	  and dotted by Piersimoni et al. 2002. We also plot the period distribution for M~3 
      \citep[full line][]{cc2001}, but these data have been divided by a factor two, in order to
	  allow an easy comparison with NGC~3201. In  spite of the different total numbers, 
	  the distributions are very  similar. On the right we plot the NGC~3201 
	  period distribution by Piersimoni et al. (dotted) and its simulation (full line), 
	  discussed in the text.
            }
         \label{fig3}
   \end{figure*}
   
However, there are two important facts to be mentioned: i) the detailed colour
distribution along the HB is by no means uniform, and ii) the RR Lyr period
distribution appears strongly peaked, a feature that cannot be understood in
terms of a more or less uniform mass distribution \citep{castorn1981, rood1989,catelan2004,
castellani2005}.  We have examined in detail this case \citep{caloidantona2008}; since there are
several other clusters showing the same problem, we summarize the main points
of the model for M3 and extend the interpretation to other clusters.

\begin{figure*}
   \centering
   \includegraphics[width=8cm]{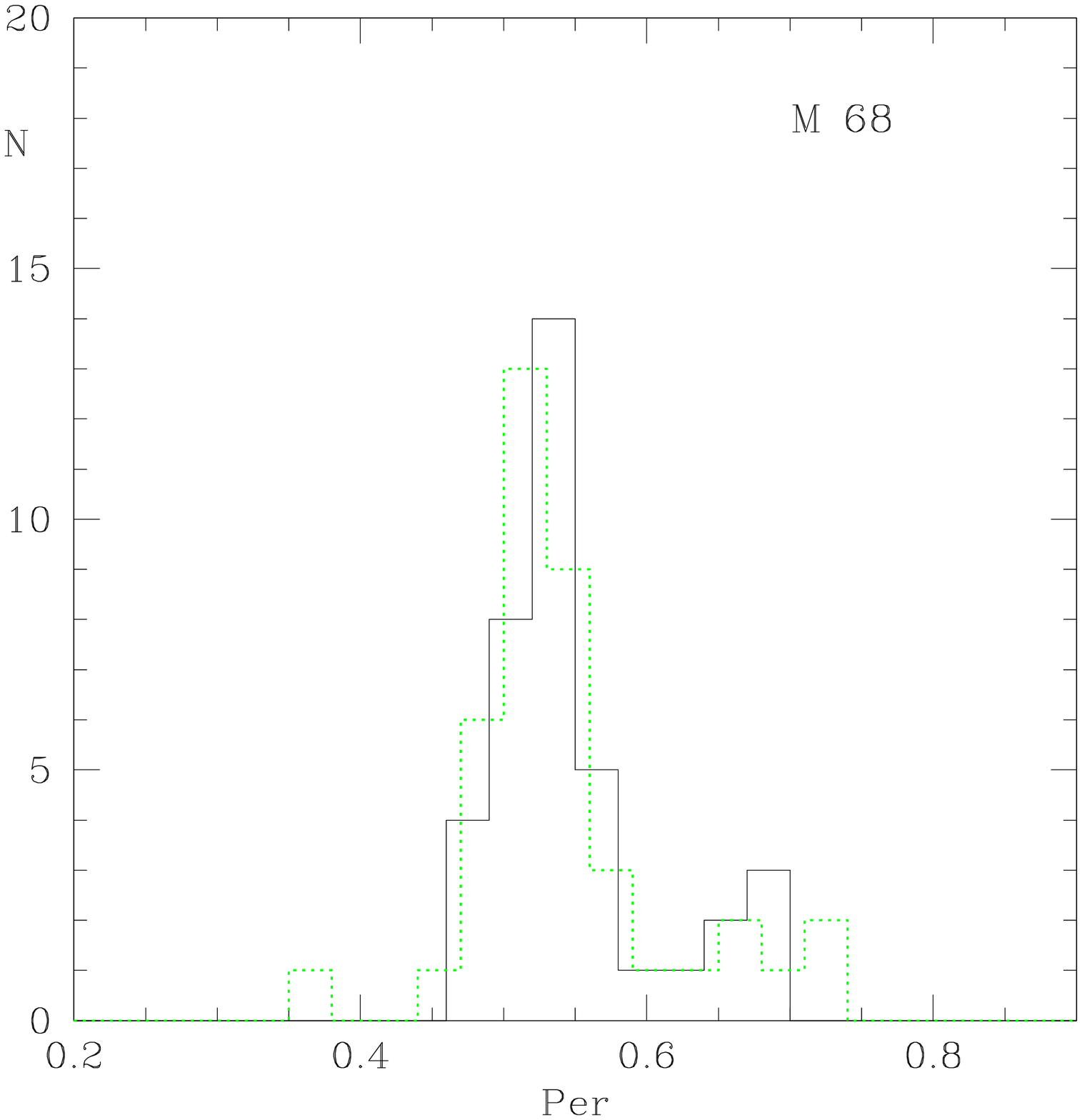}
   \includegraphics[width=8cm]{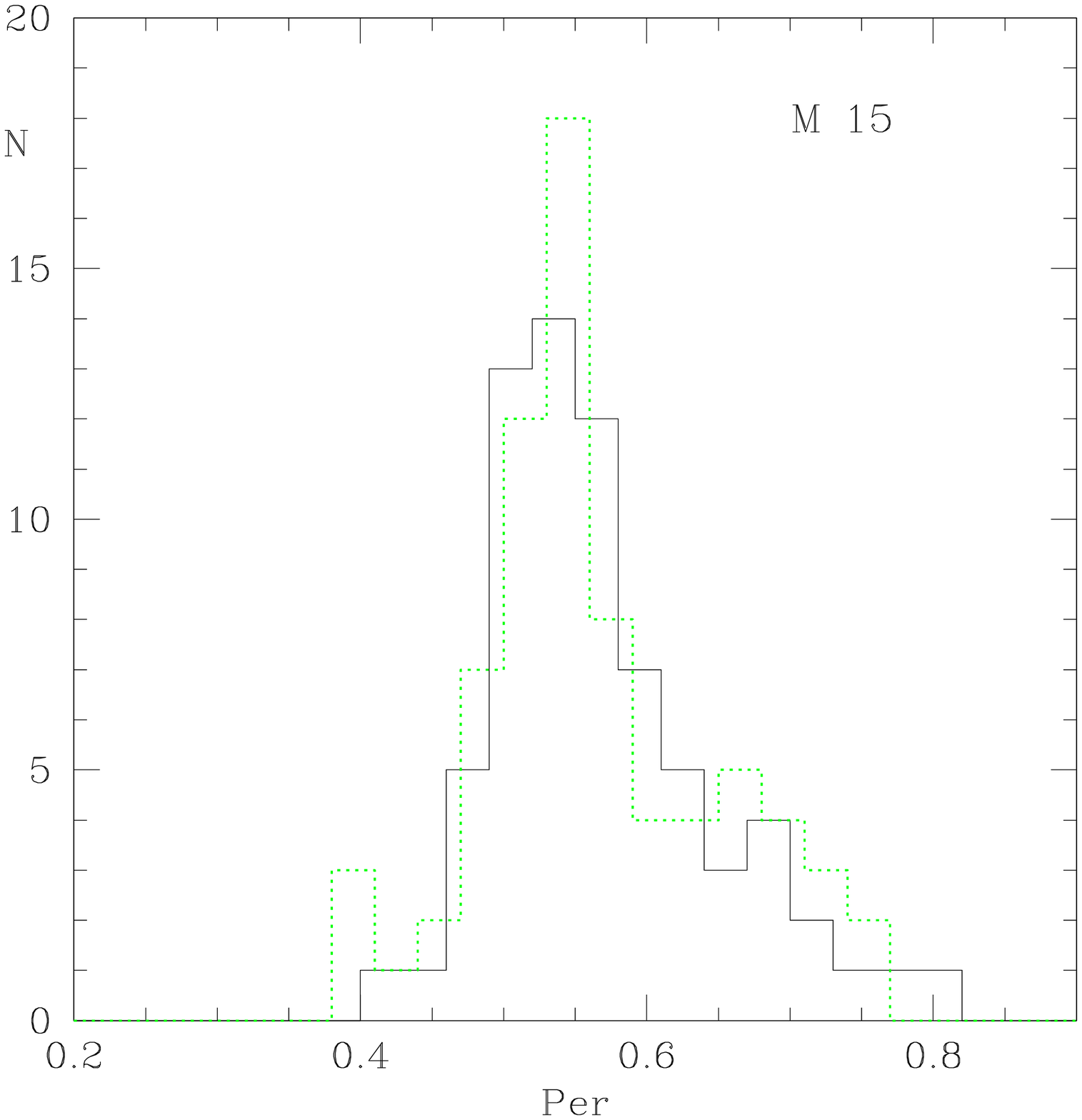}
      \caption{Period distribution fits for M~68 (left) and M~15 (right). Dotted histograms
show the observed data, full line histograms are the simulations.}
         \label{fig4}
   \end{figure*}   
   
\subsection{The period distribution of M~3}
As said before, the RR Lyr period distribution in M~3 is highly peaked (see 
Figure 3, left side, full line histogram). \cite{castellani2005} realized that 
the only way to reproduce this peak was to reduce the dispersion in the mass 
lost in the RGB. Once obtained the correct distribution with a given recipe for the 
mass loss, the authors had to assume a different average mass loss, with a 
different dispersion, to account for the blue side of the HB. In the hypothesis 
of multiple helium enhancements, the blue side is naturally populated by helium 
rich stars, as shown by \cite{caloidantona2008}. In the simulations, we can 
explain both the period distribution and the colour distribution along the HB 
for the red, variable and blue regions. The analysis however poses another problem: we 
find that the dispersion in mass loss along the RGB must be at 
most $\sigma \sim 0.003$\Msun\ to be consistent with the period distribution. 
The question remains whether this small dispersion is peculiar to M~3, or 
rather we have always been misled by the overall reproduction of the HB 
morphology, for which previous simulations (at constant Y) needed a dispersion in mass loss of 
some hundreths of \Msun, as quoted before.

\subsection{NGC 3201, NGC~7006, M5}
We examined NGC~3201 according to the same scheme used for M~3, on the basis of 
the data by \cite{layden2003} and \cite{piersimoni2002}. Unfortunately, the 
cluster is affected by differential reddening  \citep[][ and references 
therein]{vbmateo2001}. While it is possible to perform reddening estimates for 
the single RR Lyrae variables (Sturch 1966, Blanco 1992; see Layden \& 
Sarajedini 2003), the same is not so easy for non variable stars. In fact, when 
we tried to reconstruct the colour distribution on the HB, we found many non 
variable stars in the variable region. Therefore we did not attempt to 
reproduce the detailed colour distribution as we did for M~3, but only the RR 
Lyr period distribution and the overall division among red, variable and blue 
members.

We compared the period distributions in M~3 and in
NGC~3201, using both the data by Piersimoni et al. and Layden \&
Sarajedini. Notwithstanding the large difference in the total numbers of
variables with an established period (more than 200 variables in M~3, slightly
more than 50 in NGC~3201), one finds a strong similarity in the period
distributions (see Fig. \ref{fig3}, left side). So we have again a peaked distribution,
and even a dip at the same period! Therefore the NGC~3201 
solution is quite similar to the one for M~3. Our best simulations give 37\% of
HB members with helium enhanced from Y=0.26 to 0.28, that is, 
coincident with the percentage of the blue HB stars. The mass dispersion optimum
is even smaller than in M3 ($\sigma \sim$ 0.0015 \msun). The period simulation
is shown (full line histogram) in the right side of Fig. \ref{fig3}.

What said for NGC~3201 can be repeated for NGC~7006, whose RR Lyrs show again
a peaked period distribution \citep{wehlau1999}. Given the morphology of the
HB, composed mainly by red and variable stars, the required percent of helium
enhanced objects is of about 27\%, up to Y about 0.27.

The case of M~5 (NGC~5904) appears less simple, since the RR Lyrs period
distribution presents two peaks. We can only approximate this distribution,
while we succeed in reproducing the detailed colour distribution along the HB 
given by \cite{sandquist2004}. Within these limits, we estimate 70\% of
SG stars, up to a helium content of about 0.31 (we remind that M5 has more
blue HB members than M~3, with the peak of population in the blue). 

\subsection{Very metal poor clusters: M~15 and M~68}

A very good fit of the RR Lyr peaked period distribution in M~15 and
M~68 is obtained with the 
same method adopted for M~3, but obviously using the tracks of metallicity 
Z=2$\times 10 ^{-4}$, adequate to describe these two clusters. The resulting 
period distributions are in Fig. \ref{fig4}.  We must say that the fit of HR 
diagram, on the contrary, is much less successful than for the other clusters 
we have analyzed so far. The discrepancy is the following: the blue part of the 
HB in the simulations is achieved with Y in the range 0.26--0.28 for M~68, and 
0.26--0.30 for M~15, but the luminosity of the blue side is too large with respect 
to the data. We suggest that helium is not the only parameter varying in these clusters,
and that also the total CNO abundance should be varied, giving  origin to
a more complex description, still to be explored (see, e.g., the problem of NGC~1851
described above). 

\section{The ``Blue--HB" clusters }

 A specially intriguing case is presented by the clusters with a prevalently 
 blue HB, that is, with a HB type 1 -- 2  
 \citep{dickens1972}. Almost -- if not all -- of them fall in 
 the category of the classical ``second parameter'' problem. A  fact that is 
 not always considered as it deserves is that these clusters  represent {\it 
 the majority} of the population of the intermediate metallicity GCs 
 \citep[][]{alcaino1999}. So these clusters are not an exception, but rather 
 the rule, that is, the most common result of the GC formation process. We 
 shall examine some of these cases.

\subsection{M~13: multiple populations, but all belonging the SG?}

We analysed in detail the case of M~13, finding substantial support to the 
hypothesis that only the second star generation has survived in the cluster. 
This conclusion was reached on the basis of the relative positions of the turn--off, 
the red giant bump and the horizontal branch \citep{caloi-dantona2005}. 
This result, if confirmed, would be the first direct evidence for the presence 
of a substantial helium enhancement in GC stars.  In the Appendix B we present 
a detailed analysis of the HB of this cluster, in the HST plane F555 vs. F336-
F555 from \cite{ferraro1998}. The helium distribution is compared with the one 
necessary to reproduce the blue HB of NGC~2808, with the same assumption on 
age, metallicity and mass loss on the RGB. The comparison shows that M~13 stars have a
shallower distribution in helium, but may have a peak at Y$\sim$0.28, and a similar peak at Y$\sim$0.38.
In addition, {\it it completely lacks the red clump stars, that is the FG stars}. 
In the \cite{ferraro1998} sample we use, there is a total of 221 HB stars, 
populating the upper, medium and low luminosity blue HB. According to the chosen
simulation, a fraction 40\%  of stars has Y=0.28-0.29, another 30\%  has 
Y$\sim$0.31--0.35, while the low HB is reproduced by taking $\sim$30\%  of stars 
at Y=0.38. Thus the group of stars mostly contributing to the RG bump 
is that at Y=0.28--0.29, as suggested in \cite{caloi-dantona2005}. The conclusion is
that in M~13 there are multiple populations, but, according to this interpretation of the 
data, they all belong to the SG!

%INSERT REFEREE
On the other hand, several observations show the existence of chemically normal M~13 members,
for example in Na abundance \citep[e.g.][]{pilach1996} and in C and N abundances 
\citep[e.g.][]{briley2004}. However, if the Y$\sim$0.28 population is the result of
the dilution of pristine cluster matter (Y=0.24) with highly Y enriched matter (Y$\simgt$0.34),
as envisaged in many formation scenarios \citep{decressin2007,ventura2008b,dercole2008},
not necessarily the main chemical anomaly indicators will assume values noticeably
different from those of the FG stars.
%INSERT END 

The analysis of the relative positions of the turn--off, the red giant bump and 
the HB could not be performed on other blue--HB clusters, due to the lack of a 
consistent photometry for these features. Clearly a possibility is that in all 
the clusters of this group most of the first generation stars have been lost. 
There are contradictory spectroscopic indications in favour or against this hypothesis. 
In the cluster 
NGC~6752, out of nine dwarfs and nine subgiants analyzed by 
\cite{carretta2005}, only one (subgiant) has a normal nitrogen content, all the 
others are substantially N--enriched ([N/Fe]$\sim 1 - 1.7$. The recent 
new data by \cite{yong2008}, on the contrary, contain also nitrogen
normal stars. This cluster could be similar to M13, also in 
having a very helium rich population producing the extreme HB stars.

\subsection{NGC~6397, M53: apparently SSPs, but possibly SG--SSP?}

Notice that NGC~6397 has [Fe/H]=--2 \citep{gratton2001} and so is not an 
intermediate metallicity cluster. 
It has a short blue HB, lacking 
extreme HB and blue hook stars, and its HR diagram has always been regarded as a 
perfect example of SSP, especially following the exceedingly refined HST proper 
motion selected observations by \cite{king1998} and \cite{richer2006}. Nevertheless,
only three subgiants out of 14 stars 
are nitrogen normal \citep{carretta2005}, leading us to suspect
that the material from which these stars formed is CNO processed and thus of
SG. This occurrence had already been 
noticed in \cite{bonifacio2002}, with reference to the paradox that nitrogen 
rich stars had almost--normal lithium content \citep[see also][]{pasquini2008}. 

The above considerations lead naturally to the question: the simple population 
GC -- one chemical composition, one age, of which one used to speculate until 
very recently -- does it exist? We are inclined to give a negative answer. Let 
us consider the observations by \cite{liburstein2003}, who took integrated 
spectra of eight galactic GCs, ranging in metallicity from [Fe/H]$\simlt -2$\ 
to [Fe/H]$\sim -0.8$, and showing a variety of HB morphologies (the clusters 
are: M~15, M~92, M~53, M~2, M~3, M~13, M~5, M~71). All these cluster show a 
substantial N--enhancement with respect to field stars of the same metallicity. 
Since we are dealing with integrated spectra, the result can not be directly 
interpreted in terms of percentage of second generation stars, that would be 
composed by nitrogen rich (CN or CNO cycled) matter. Nevertheless, the N--
excess with respect to the field is a clear indication that {\it at least} a 
certain amount of stellar matter is not the original one, in all the eight GCs 
quoted above.

The case of M~53 presents some interesting features. It is a very metal poor 
([Fe/H] $\sim -2$,  Zinn 1985, Harris 2003), massive cluster (M$_{\rm v}= -8.70$~mag). 
Out of a total of 307 observed HB members \citep{rey1998}, 257 are located blueward 
of, 35 within, and 12 redward of the RR Lyr instability strip, the bluest stars 
being located at (B-V) $\sim$ -0.08. Three stars appear separated at bluer 
colours.  So, the HB has a very short extension in colour, with most of its 
members concentrated at (B-V) $\sim$ 0.05. This star distribution is unique 
among very metal poor GCs -- see, f.e., Table 13 in \cite{walker1994}. 
In fact: in M~15, a massive and high central density cluster, the 155 HB 
members are distributed from redward of the RR Lyrs to very blue colours and 
low visual luminosity, reaching the TO magnitude; M~68, a relatively small and 
not very concentrated cluster, exhibits a short HB from the red region to about 
(B-V) $\sim$ -0.1 mag, with almost equal numbers of blue on the one side, and 
red and variable stars, on the other.
   
The extraordinary star concentration in a small colour interval suggests that 
the largest part of the HB population in M~53 can be obtained assuming a 
dispersion in the mass loss of $\sim$ 0.01 \Msun; a more precise photometry 
could give more stringent limits to the dispersion. On this basis, M~53 would 
appear a good candidate for a ``normal'', first--generation--only cluster, but 
for the N--enhancement and the the (mild) intrinsic spread in CN bandstrength 
\citep{martell2008}. These chemical properties suggest that we are 
likely dealing with a one-generation cluster, but one in which the star 
generation we observe is the {\it second} one, and not the first.

Of course a final answer to the initial question will have to wait the 
investigation of the whole body of Galactic GCs, but since now secondary 
episodes of star formation appear widespread and crucial for building--up the 
clusters themselves.
 
\section{Conclusions: how did the GCs form?}

In all GCs examined in this work, a large fraction of the stellar population 
takes origin from secondary star formation episodes. Notice that we have examined 
only a fraction of the clusters with HB morphology or RR Lyr period distributions similar to those
described here, so that we can suggest that the results of this work probably hold
for a larger population of Galactic GCs. While the most massive 
clusters have extreme helium enhancements, also moderately massive clusters 
show a considerable degree of helium variation. 

We reached our goals by examining in detail GCs that have unexplained features in their HBs, 
and extending the results to clusters with similar features. 
The HB morphology is one of the important
features: clusters having a bimodal or multimodal HB are most easily interpreted by 
the coexistence of multiple generations with different helium content. 
Also a unique SG, whose 
stars had different degrees of mixing with pristine matter and thus ended up with different helium
contents, is a possible solution. In any case, the extreme helium rich populations 
(in \ocen\ and
NGC~2808 at least) are neatly separated from the other MS stars so that they should have 
a well defined independent origin.  

Further, we used the period distribution of
RR Lyrs in several clusters in order to reject the hypothesis of a unique Y value 
with a relatively large
spread of mass loss on the RGB, that has been the standard way of interpreting the whole HB colour
extension, but is inconsistent with most period distributions.

We re--examined in detail the HB distribution in NGC~2808, and obtained a helium distribution
consistent with the main sequence recent data by \cite{piotto2007}. Also the UV data
of this cluster find a good interpretation in terms of population with varying helium content,
if we make the further hypothesis of deep mixing to understand the location of the blue hook 
stars. We also show that simulation of M~13 UV data is well explained with populations having
different helium.

After this analysis, then, we must face the problem that the SG formation is not a peculiarity of
a few very massive clusters, but must be the normal way in which a GC is formed in our Galaxy.
It is almost obvious, and has often been discussed in the literature, that the 
ejecta of a unique first stellar generation with a normal initial mass function 
(IMF) can not produce enough mass to give origin to such a large fraction of 
second generation stars (see, e.g., the case made by Bekki and Norris (2004), 
for the blue main sequence of $\omega$Cen).
The only solution to the IMF problem is that {\it the starting initial mass from which 
the first generation is born was much larger than today first generation 
remnant mass} (at least a factor 10 to 20 larger), so that the processed ejecta 
of the first generation provide enough mass  to build up the second one. 
There are two possible ways of producing this result: 
the first possibility is that all these GCs formed
within a dwarf galaxy environment \citep{bekki2006,bekki2007}. There, GCs may be formed 
by mixing of pristine gas with the winds of the very numerous massive AGB stars 
evolving in the field of the dwarf galaxy, 
and later on the dwarf galaxy is dynamically destroyed.
A second possibility has been recently suggested by \cite{dercole2008}.
They  show that the SG stars are preferentially born in the inner core 
of a FG cluster, where a cooling flow collects the gas 
lost by the FG stars. The massive stars that explode as SN~II were 
preferentially concentrated in the cluster core. After the mass loss due to the 
supernovae type II explosions, the cluster expands, and begins losing the stars ---mainly
of FG--- going out of the tidal radius. Thus the cluster may
be destroyed, unless the gas lost by the most massive AGB  
stars begins collecting in the core and forms the SG, that initially does not take part in the
cluster expansion. 
The study of the cluster dynamical evolution, followed by means of
N-body simulations, shows that the cluster preferentially loses FG
stars; these simulations show that high SG/FG number ratio can be
achieved and SG-dominated clusters may survive.

\section{Acknowledgments} 
Many colleagues contributed to this analysis in several ways: we thank G. Bono
for the HST UV data for NGC~2808, and F. Ferraro for the M~13 HST data. 
S. Cassisi helped by providing the colour transformations. Conversations and a long
collaboration with A. D'Ercole and E. Vesperini convinced one of us (FD) that the
SG is a necessary ingredient for the survival of globular clusters.

\begin{appendix}
\section{A new analysis of NGC~2808 optical and ultraviolet data}
  
   \begin{figure}
   \centering
   \includegraphics[width=7cm]{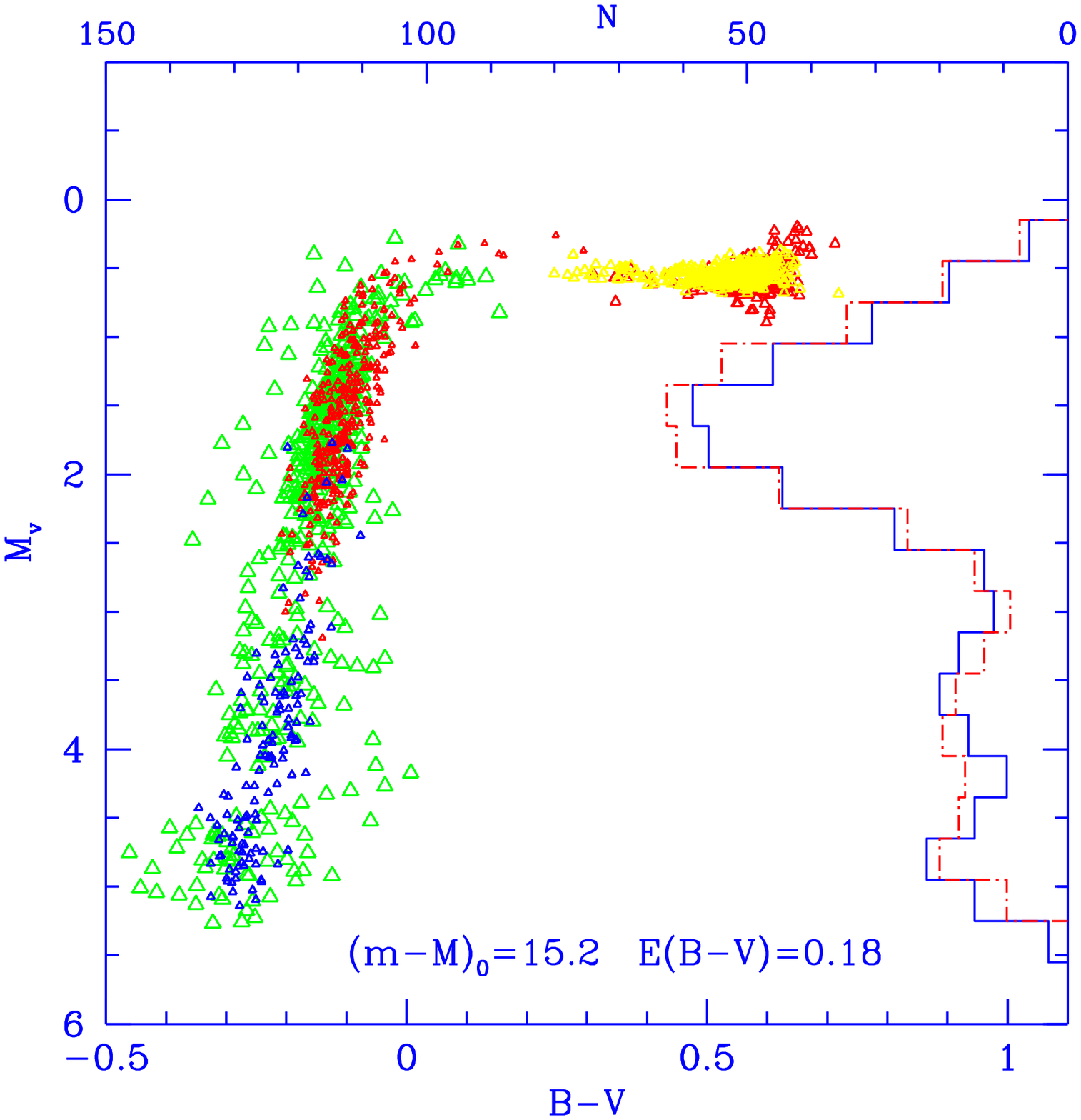}
      \caption{The HB data by Bedin et al. (2000) (open triangles) and their simulation 
	  superimposed (full triangles). The histogram of the blue HB data is shown as a full histogram
	  on the right, and the simulated histogram is dot--dashed. The lower luminosity
	  clumps EBT2 and EBT3 are both obtained with a unique value of Y=0.385, as explained
	  in the text. }
         \label{figA1}
   \end{figure}

   \begin{figure}
   \centering
   \includegraphics[width=7cm]{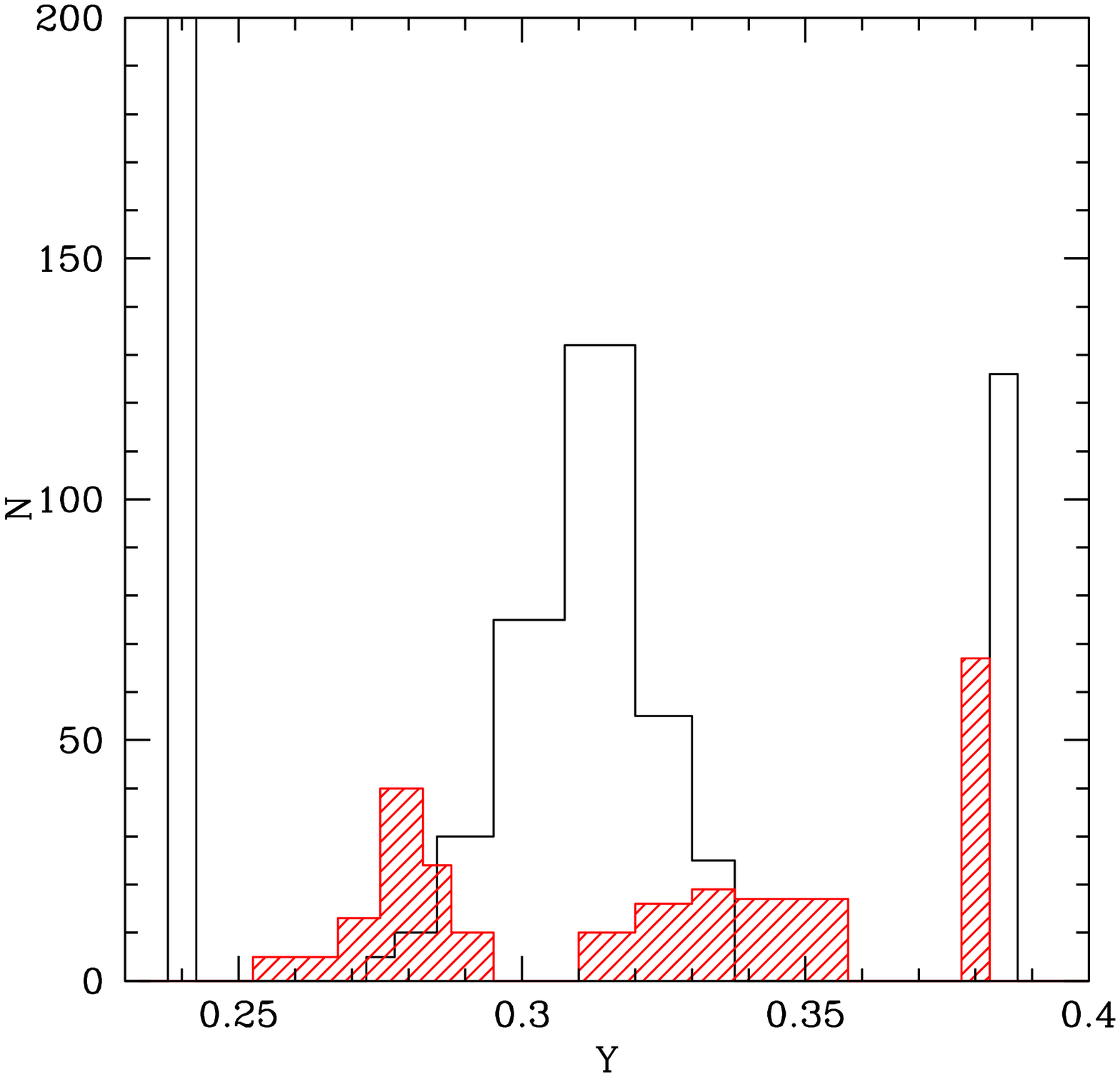}
      \caption{The full histogram represents the number vs. helium content distribution of
	  the NGC~2808 HB of Fig. A1. The dashed histogram represents the corresponding distribution 
	  for M~13. Notice that the M~13 distribution completely lacks stars with normal Y=0.24.}
         \label{figA2}
   \end{figure}
\subsection{The V vs. B--V data}

  \begin{figure*}
   \centering
      \includegraphics[width=5.2cm]{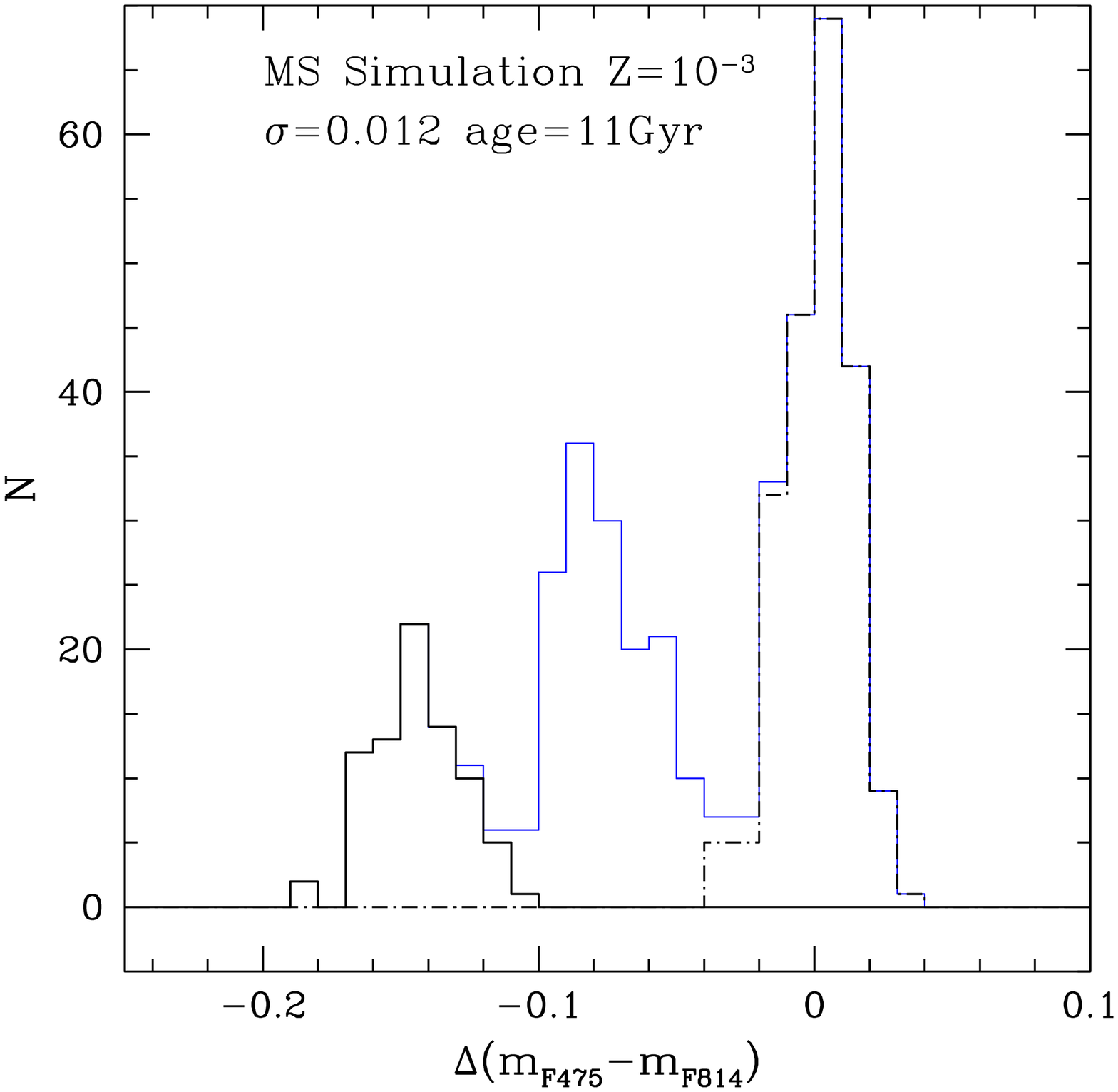}
   \includegraphics[width=5.2cm]{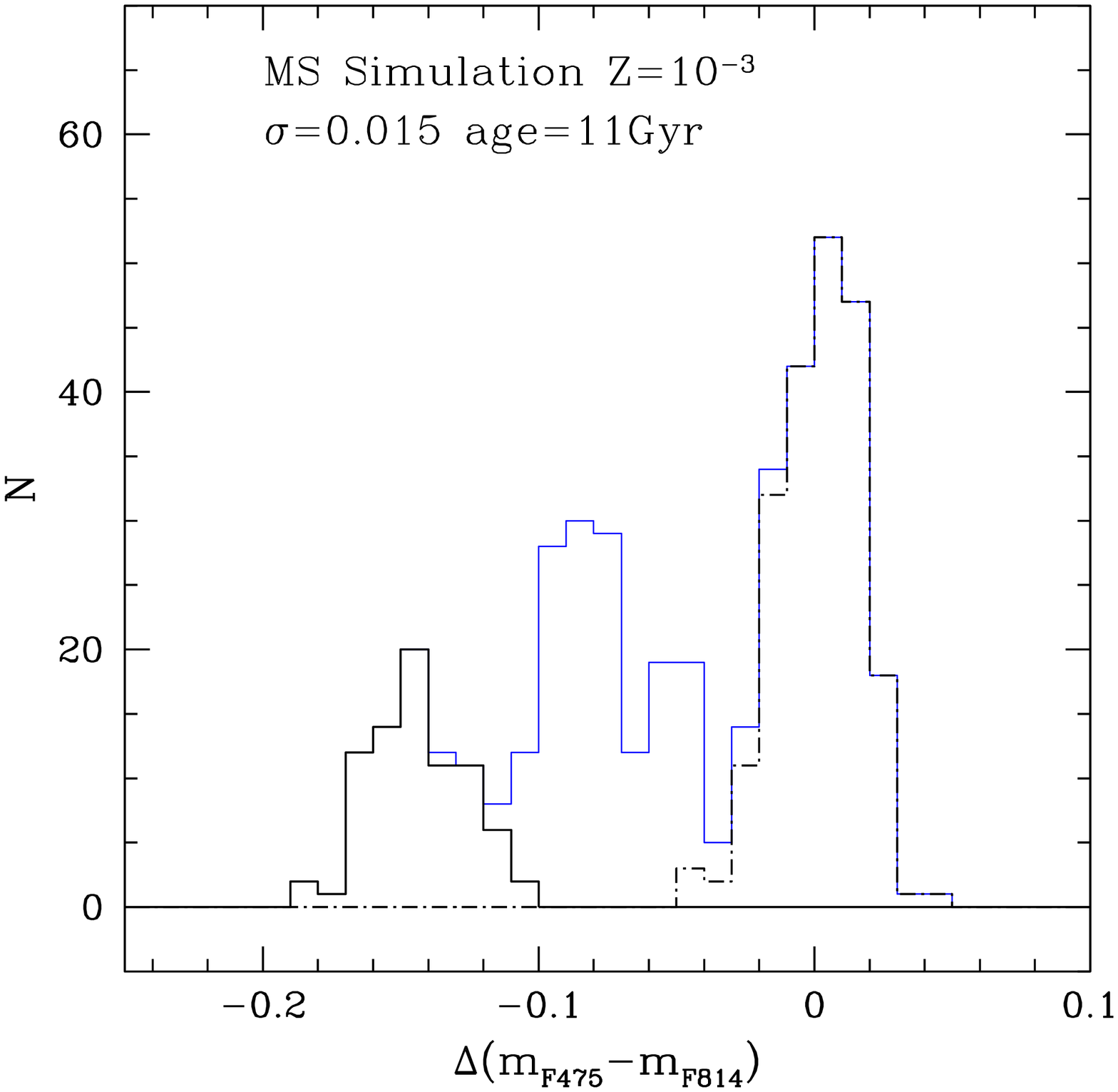}
   \includegraphics[width=5.2cm]{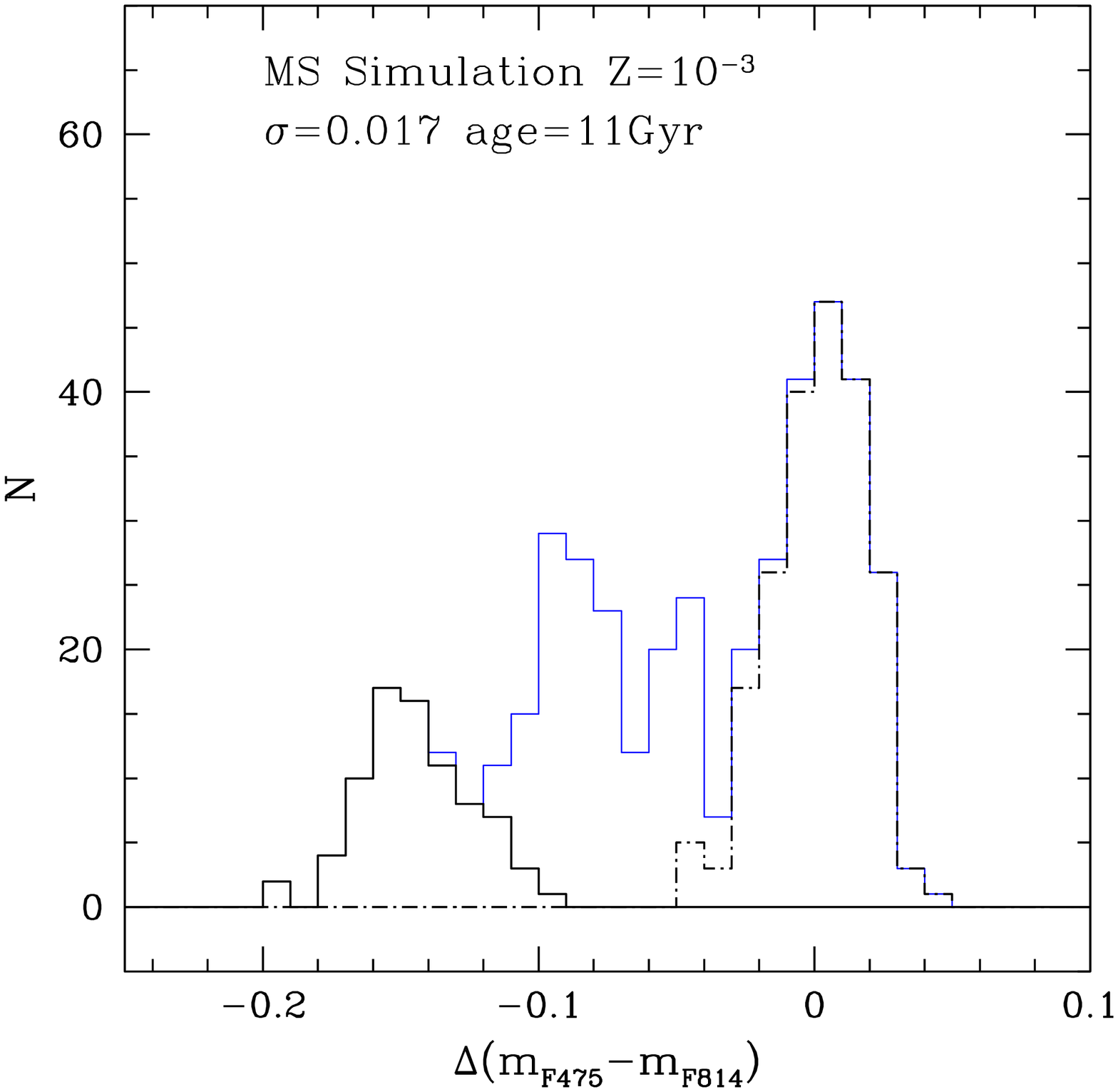}
      \caption{The figures show simulation of the main sequence colour distribution corresponding
	  to the N(Y) distribution of Fig. \ref{figA2} in the ACS colour F475-F814 of Piotto et al. 2007.
	  The simulation is shown for the magnitude interval 5.0$\leq M_{814}\leq$6.0.
	   The three panels correspond to different assumed errors 
	   on the colour (0.012, 0.015 and 0.017mag).}
         \label{figA3}
   \end{figure*}

We analyzed the \cite{bedin2000} data of NGC~2808 twice 
\citep{dantonacaloi2004, dantona2005}, but always assuming that the mass lost 
on the RGB has a slight but positive dependence on the helium content of the 
sample. Having now shown that this is not the case (Sect. \ref{sec:2}) it is 
reasonable to make another analysis, and derive the N(Y) distribution that fits 
the HB, obtained by assuming that both $\delta M_0$\ and $\sigma$\ do not 
depend on Y. For the simulation, we closely follow the procedure by 
\cite{dantonacaloi2004} and \cite{dantona2005}, apart from the analysis of the 
blue hook stars \citep[the clump EBT3 in the definition by ][]{bedin2004} for 
which we make the further assumtions described  below. The problem is that we 
{\it can not adopt a different helium content for each of the clumps EBT2 and 
EBT3}, because, using the N(Y) distribution that reproduces the HB, we have to 
reproduce also the colour distribution in the MS data by 
\cite{piotto2007}. For this aim, we convert our main sequence stellar models 
into the ACS bands F475 and F814 by means of the transformations provided by 
\cite{bedincastelli2005}, and simulate the MS colour distribution.  We can not 
make a detailed comparison with the data, as they are not available to us, but 
simply show the histogram of number versus the colour difference (in F475--F814) 
from the reference main sequence of Y=0.24, in the magnitude interval $5 \leq 
M_{814}\leq 6$. Comparison with the histograms in Figure 3 in \cite{piotto2007} 
shows that there is a fair reproduction of the observations, if the colour error is 
taken to be 0.017mag (indicated by $\sigma$ in the top of Fig. \ref{figA3}). 
\begin{figure*}
   \centering
   \includegraphics[width=8cm]{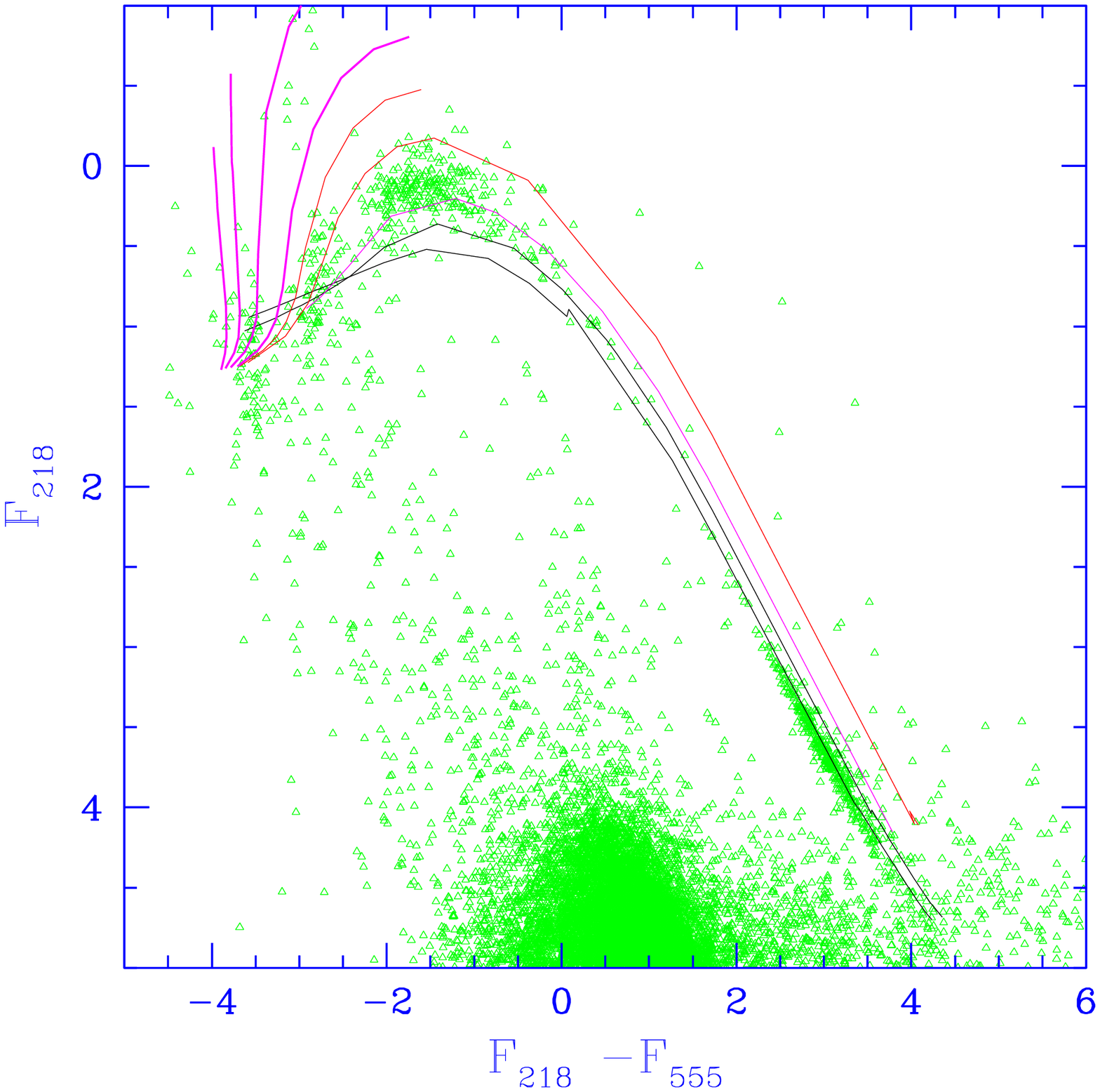}
   \includegraphics[width=8cm]{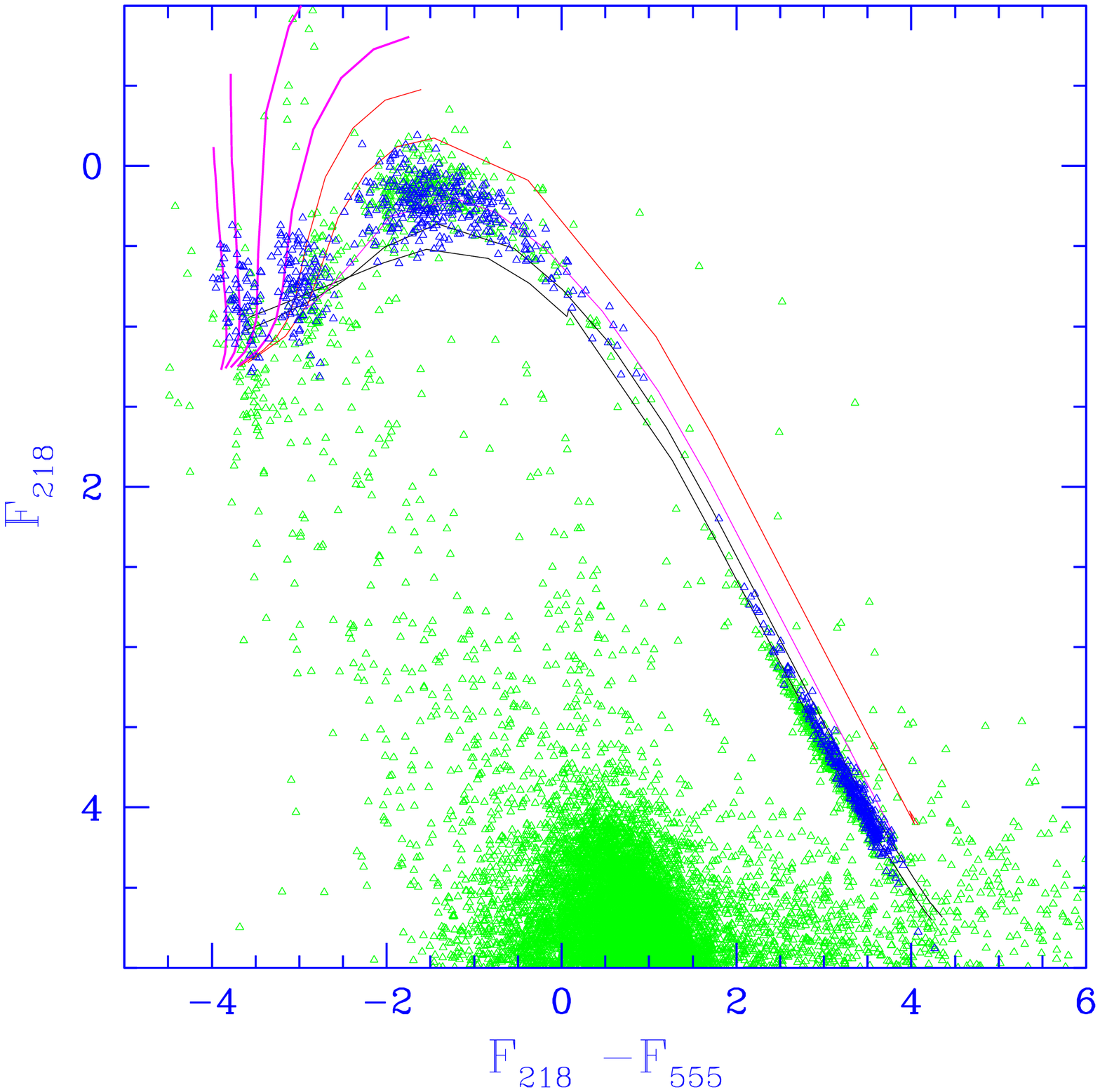} 
       \caption{We show the ZAHBs of Z=0.001 with different helium content in the 
	   absolute UV magnitude F218 versus F218-F555 colour.
	   The Y=0.24 ZAHB line is used to define the distance modulus and reddening
       of the dataset by Castellani et al. 2005. We impose that the red clump data (on the right
	   bottom part of the left figure) lie symmetrically on this ZAHB. 
	   The rising lines on the left are ZAHBs of models with the same core
	   mass of the Y=0.40 sequence, but having larger envelope helium abundance. From left to 
	   right, we have Y=0.8, 0.7, 0.6, 0.5 and 0.45. On the right, we show the simulation
	   of Fig. A1.  }
         \label{figA4}
   \end{figure*}   

Let us discuss in detail the assumptions made to fit the EBT2 and EBT3 clumps. 
As the blue MS is well separated from the other stars, all the stars in this 
blue MS share the same helium abundance. We then take a unique high helium 
abundance for all the stars in the clumps EBT2 and EBT3  
\citep[as we assumed also in][]{dantona2005}. Attributing to the cluster an age of 11Gyr, the average 
mass loss rate necessary to fit the red clump is 0.18\Msun. Due to the spread 
in mass loss ($\sigma$=0.008\Msun) assumed in order to fit the width of the red 
clump, and due to the choice of a very high helium in order to reproduce the 
blue MS, the remnant mass is in a very strict range close to the helium flash mass 
(M$\sim$0.49\Msun, to be compared with M$_c$=0.4676\Msun, for Y=0.40). We will 
fix a value \mlate, above which we put the star on the track corresponding to 
its mass and helium content. Below \mlate, we assume that the star has suffered 
very deep mixing, and that {\it its helium surface abundance has increased}. We 
parametrize the resulting surface helium abundance between fixed values, and 
distribute random the stars along the corresponding tracks. By this hypothesis, 
we are able to reproduce well the gap between EBT2 and EBT3 {\it without 
invoking a helium discontinuity}, as shown in Fig. \ref{figA1}. We summarize the parameters 
chosen for this fit: age of 11Gyr, $\Delta M_0=0.18$\Msun, $\sigma=0.008$, 
\mlate=0.487\Msun, and the N(Y) is shown in Fig. \ref{figA2}.
   
\subsection{The HST UV data of NGC~2808}
The optical bands are certainly not the best to describe the very hot HB stars in NGC~2808. The
HST data have shown that these objects are the most 
luminous ones in the UV bands \citep{brown2001}. \cite{lee2005} first attempted to fit the
data with HB models having varying helium content. Here we use the data by \cite{castellani2006}
in the plane F218 versus the colour F218-F555. Our models have been transformed into these
bands by using the \cite{origlia2000} colour transformations. Fig.~\ref{figB1} shows on the left
the data and the ZAHBs. The choice of the distance modulus and reddening in the colour F218-F555
have been made in order to fit the red clump data on the Y=0.24 ZAHB. This choice
leads to a good superposition of the highest luminosity stars \citep[corresponding
to the EBT1 clump by][]{bedin2004} above the ZAHB of Y=0.32.
The rising short lines on the left are ZAHBs of models with the same core
mass of the Y=0.40 sequence, but having larger envelope helium abundance. From left to 
right, we have Y=0.8, 0.7, 0.6, 0.5 and 0.45. The simulation shown in Fig \ref{figA1} for the B and V 
colours is shown in the right part of Fig. \ref{figA4} in these UV bands. The fit of red clump, EBT1 and EBT2 is 
very good, showing once for all that there is a helium difference between the red clump, EBT1 and EBT2.
The gap between EBT2 and EBT3 results well simulated by assuming that a fraction of
the stars born with Y=0.40 suffers deep mixing, increasing its Y to the range 
Y=0.7--0.8. Nevertheless, the simulation fails to reproduce the lowest luminosity EBT3 stars.
Reasons for this result are several: first of all, these stars have atmospheres which are not only
very helium rich as we assumed, but also carbon rich, and the \Teff--colour transformations
can not take this into account. Second, it is well possible that these stars ignite helium with a late 
flash at a core mass smaller than we assumed, and thus have a smaller intrinsic luminosity.

\section{A possible fit for the HB of M~13}
\begin{figure}
   \centering
   \includegraphics[width=7cm]{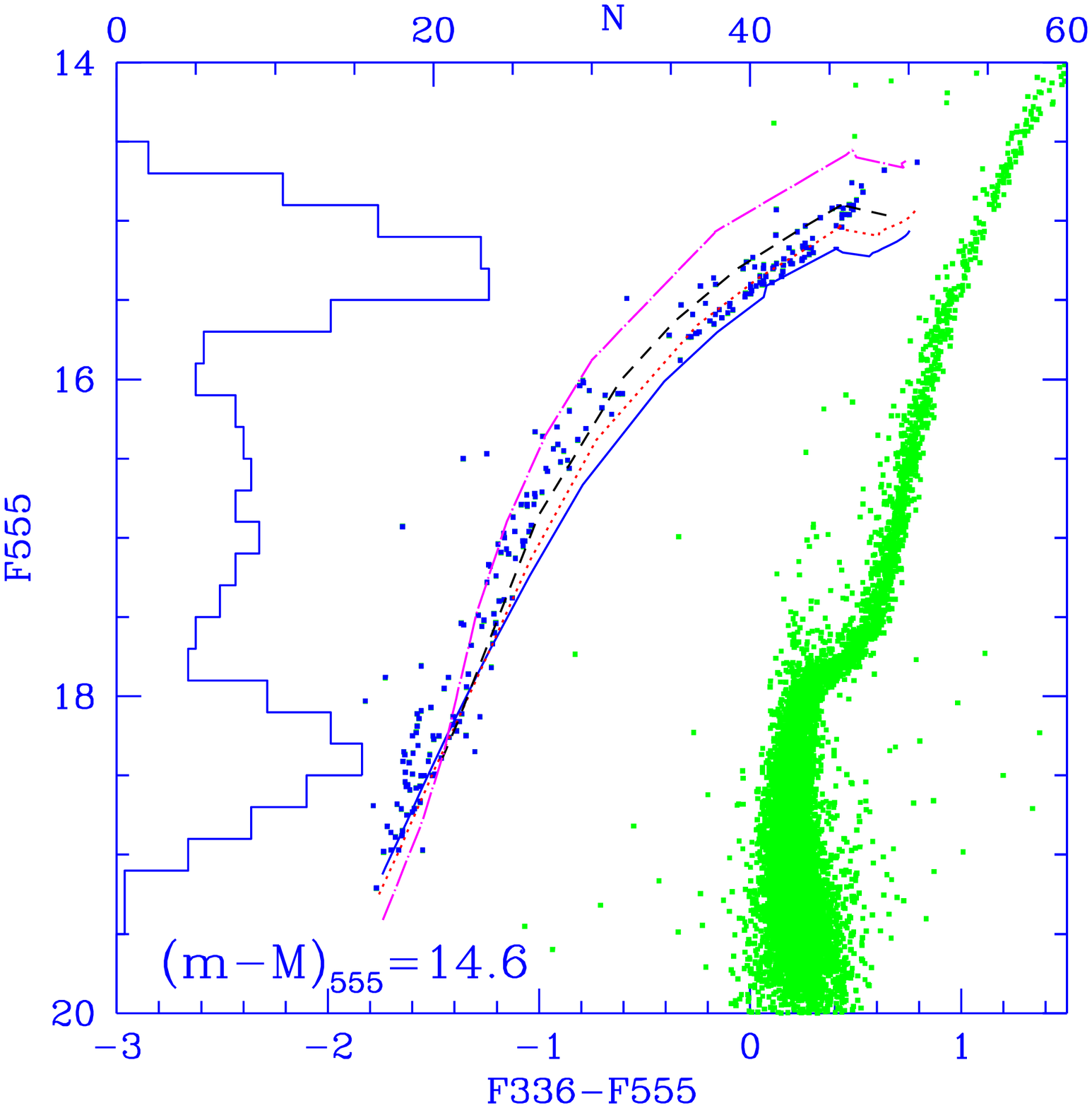}
       \caption{ M~13 HB data by Ferraro et al.(1998) compared with our ZAHBs in the plane
	   F555 vs. F336-F555. The distance modulus is chosen so that the Y=0.28 ZAHB (dotted line)
	   fits the upper luminosity clump. The other lines are the 
	   ZAHBs for Y=0.24 (full line), Y=0.32 (dashed) and Y=0.40 (dash dotted).
	   On the left we show the histogram of HB counts as a function of F555 magnitude.}
         \label{figB1}
   \end{figure}   
\begin{figure}
   \centering
   \includegraphics[width=7cm]{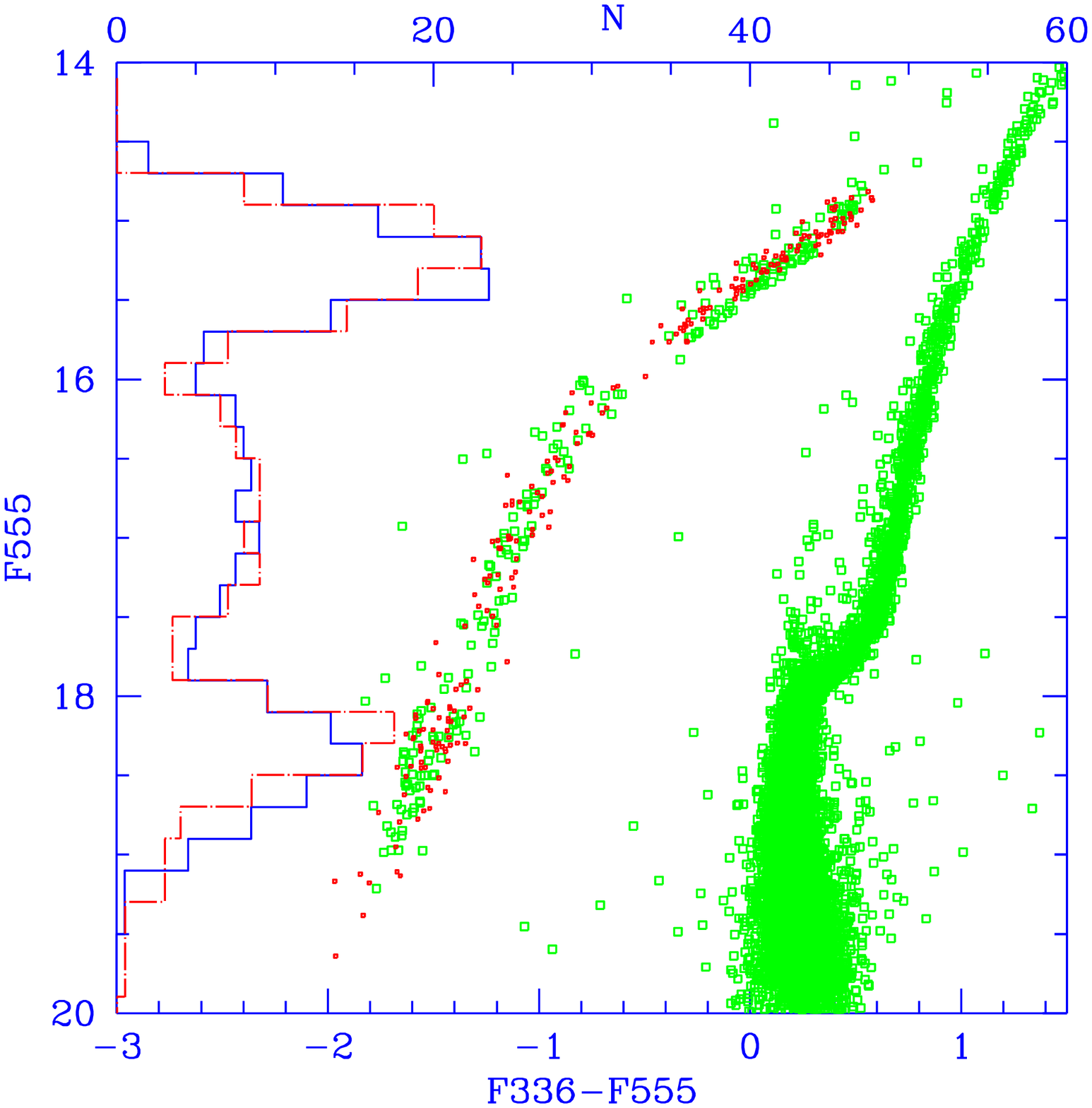} 
       \caption{Superimposed to the data, we show the
	   simulation done with the same values, $\delta$M=0.18\Msun\ and age 11~Gyr, 
	   chosen for the NGC~2808 simulation. $\sigma$ is 0.01\Msun\ and \mlate=.479\Msun
	   }
         \label{figC2}
   \end{figure}   

\cite{caloi-dantona2005} provocatively proposed that the blue--HB clusters, 
and in particular M~13, have 
completely lost their FG (see Sect. 6.1). As the metallicity of M~13 and NGC~2808 are close,
and the blue HB of the two clusters are morphologically similar, we try to fit the HB of
this cluster by imposing a simulation in which the age and average mass loss are the same as
in the NGC~2808 fit, but {\it the red clump population, at Y=0.24, is totally eliminated}. 
Fig.\ref{figB1} shows the HST data by \cite{ferraro1998} in the plane F555
vs. F336-F555. We superimpose our ZAHBs as in Fig.~\ref{figB1}, assuming a visual distance
modulus of 14.6mag and zero reddening. The modulus is chosen {\it so that the Y=0.28 ZAHB
coincides with the upper clump}, in agreement with the result by \cite{caloi-dantona2005} that 
attributed Y=0.28 to the dominant cluster population. We see that, with this choice, the
ZAHB of the middle HB is again consistent with Y$\sim$0.32. This difficulty of fitting the HB with
a unique ZAHB had already been pointed out by \cite{grundahl1998} in their analysis of 
the HR diagram of M~13 in the Str\"omgren colours, and signalled also for the clusters 
NGC 288 and NGC 6752. These authors indeed attributed this unexpected feature to the presence
in M~13 of two distinct HB populations, one of which had undegone deep mixing, following 
\cite{sweigart1997}. In our interpretation, these stars have a higher helium content already starting
from their formation.

Fig. \ref{figC2} shows the HB simulation superimposed to the M~13 data. 
Apart from the lack of the Y=0.24 part, the distribution N(Y) is different indeed from 
that obtained for NGC~2808, as we show in Fig. \ref{figA2}, but the lowest part of the HB
can be interpreted again as a very  high helium population (Y=0.38). Notice however, that M~13
does not contain the large population of blue hook stars present in NGC~2808, and this 
difference remains to be explained.
 
\end{appendix}

\label{lastpage}

\end{document}